\documentclass[12pt]{article}
\usepackage{authblk}

\usepackage[letterpaper,top=2cm,bottom=2cm,left=2.5cm,right=2.5cm,marginparwidth=1.75cm]{geometry}

\usepackage{setspace}

\usepackage{amssymb}
\usepackage{graphicx}
\usepackage{amsmath,amsfonts,amssymb,breqn}
\usepackage[TS1,T1]{fontenc}
\usepackage{textcomp}
\usepackage{graphicx}
\usepackage{caption}
\usepackage{subfig}

\usepackage{float}
\usepackage{booktabs}

\usepackage[colorlinks=true, allcolors=blue]{hyperref}

\usepackage{tikz}
\usetikzlibrary{matrix,calc}
\usepackage{epstopdf}

\usepackage{subfig}

\usepackage{algorithm}      
\usepackage{algpseudocode}

\title{HWF-PIKAN: A Multi-Resolution Hybrid Wavelet-Fourier Physics-Informed Kolmogorov-Arnold Network for solving Collisionless Boltzmann Equation}

\author[1]{Mohammad E. Heravifard\footnote{PhD Candidate}}
\author[1]{Kazem Hejranfar\footnote{Professor}\thanks{Corresponding author: khejran@sharif.edu (K. Hejranfar)}}
\affil[1]{\it{Department of Aerospace Engineering, Sharif University of Technology, Tehran, Iran}}
 
\date{}

\begin{document}
\maketitle	
\begin{abstract}
Physics-Informed Neural Networks (PINNs) and more recently Physics-Informed Kolmogorov–Arnold Networks (PIKANs) have emerged as promising approaches for solving partial differential equations (PDEs) without reliance on extensive labeled data. In this work, we propose a novel multi-resolution Hybrid Wavelet-Fourier-Enhanced Physics-Informed Kolmogorov–Arnold Network (HWF-PIKAN) for solving advection problems based on collisionless Boltzmann equation (CBE) with both continuous and discontinuous initial conditions. To validate the effectiveness of the proposed model, we conduct systematic benchmarks on classical advection equations in one and two dimensions. These tests demonstrate the model's ability to accurately capture smooth and abrupt features. We then extend the application of HWF-PIKAN to the high-dimensional phase-space setting by solving the CBE in a continuous-velocity manner. This leverages the Hamiltonian concept of phase-space dynamics to model the statistical behavior of particles in a collisionless system, where advection governs the evolution of a probability distribution function or number density. Comparative analysis against Vanilla PINN, Vanilla PIKAN, as well as Fourier-enhanced and Wavelet-enhanced PIKAN variants, shows that the proposed hybrid model significantly improves solution accuracy and convergence speed. This study highlights the power of multi-resolution spectral feature embeddings in advancing physics-informed deep learning frameworks for complex kinetic equations in both space-time and phase-space.

\end{abstract}
	

\newpage

\section{Introduction}

In recent years, intelligent learning-based approaches have expanded beyond data science into computational science, giving rise to scientific machine learning (SciML), a promising suite of tools for challenging problems in scientific computing. In computational fluid dynamics (CFD), traditional discretization-based schemes remain widely used but still face difficulties such as grid generation for complex geometries, discretization-induced numerical errors, and challenges related to nonlinearity \cite{challenge,CFDML}. Unlike purely data-driven models, SciML enables the integration of governing partial differential equations (PDEs) directly into learning frameworks. Artificial neural networks (ANNs), supported by the universal approximation theorem and the Kolmogorov–Arnold representation theorem, are particularly suitable for this paradigm due to their expressiveness and differentiability. Early work by Lagaris et al. \cite{lag} introduced the use of ANNs for solving ODEs and PDEs, while previous ANN applications focused mainly on algebraic equations \cite{wang,lee}. This line of research was revitalized by Raissi, Karniadakis, and collaborators \cite{raisi} through physics-informed neural networks (PINNs), which leverage automatic differentiation to solve forward and inverse problems involving linear and nonlinear PDEs. PINNs have since been applied to various fluid-flow problems, including incompressible and compressible Navier-Stokes equations \cite{NSF,raisi4,highspeed}, aerodynamic flows around cylinders and airfoils \cite{cylinder,incom}, and internal flows, often achieving competitive efficiency relative to traditional CFD. More recently, Kolmogorov-Arnold Networks (KANs) have further advanced PDE modeling across multiple domains. Physics-Informed KANs (PIKANs) have demonstrated improved performance in solid mechanics, multiscale physics, and dynamical systems \cite{PIKAN1,PIKAN3}, while specialized variants such as ChebPIKAN, SPIKAN, and Scaled-cPIKAN have enhanced modeling of fluid dynamics and high-dimensional PDEs \cite{PIKAN13,PIKAN14}. Additional extensions of PINNs include applications to two-dimensional phase-change phenomena \cite{phasechange}, supersonic flow over cones \cite{supersonic}, and transformer-based neural operators such as PINTO, which generalize PDE solutions under diverse initial and boundary conditions \cite{pinto}.\\

Generally, PINNs and PIKANs can be trained in either data-free or data-driven modes, yet relatively limited attention has been devoted to fully data-free formulations for solving physics-based partial differential equations. In this work, we introduce a multi-resolution Hybrid Wavelet-Fourier-Enhanced Physics-Informed Kolmogorov-Arnold Network (HWF-PIKAN), a new framework designed to improve the accuracy and robustness of data-free SciML solvers. The proposed architecture integrates global Fourier features with localized multi-resolution Ricker wavelets, enabling the network to simultaneously capture smooth large-scale structures and sharp localized variations, an ability that standard PINN/PIKAN models often lack due to spectral bias and insufficient multi-scale representation. To assess the effectiveness of the hybrid embedding, we first benchmark the proposed variants on the classical advection equation in one and two spatial dimensions. These test cases allow direct comparison against analytical solutions and illustrate how multi-resolution spectral features improve convergence and predictive fidelity. We then extend the method to the collisionless Boltzmann (Vlasov) equation, a fundamental kinetic equation governing phase-space transport in plasma physics, rarefied gas dynamics, beam propagation, and astrophysical systems. Solving the CBE is notoriously challenging due to its high dimensionality, nonlinear phase-space distortion, and the need to resolve fine-scale filamentation over long times. Traditional mesh-based schemes require expensive velocity-space discretization, as in Refs. \cite{discret,song}. In contrast, we adopt a continuous-velocity formulation that eliminates velocity discretization entirely, enabling a mesh-free representation of the full phase-space density. Within this setting, the HWF-PIKAN architecture is particularly advantageous: the global Fourier modes capture coherent transport structures, while the localized wavelets adapt to steep gradients and filamented features that naturally arise in kinetic systems. This synergy allows the proposed method to more effectively approximate solutions to CBE compared to conventional PINN/PIKAN models, especially in regimes requiring multi-scale resolution.\\

Overall, the multi-resolution HWF-PIKAN framework demonstrates strong adaptability across both low- and high-dimensional PDEs and provides a promising mechanism for modeling complex phase-space dynamics without velocity discretization. These properties position HWF-PIKAN as a valuable and broadly applicable tool for computational physics and scientific computing. The remainder of this paper is structured as follows: we first present the mathematical background of physics-informed frameworks. We then describe the algorithm and implementation of the proposed multi-resolution HWF-PIKAN and spectral feature embeddings. Next, we validate and compare the models on benchmark advection problems based upon various solvers. Finally, we extend the approach to solve the high-dimensional CBE, which is also known as Vlasov equation in phase-space and conclude with a discussion of the results.

\section{Methodology}
\begin{figure}[H]
	\centering
	\includegraphics[width=0.7\textwidth, trim=0cm 7.5cm 0cm 0cm, clip]{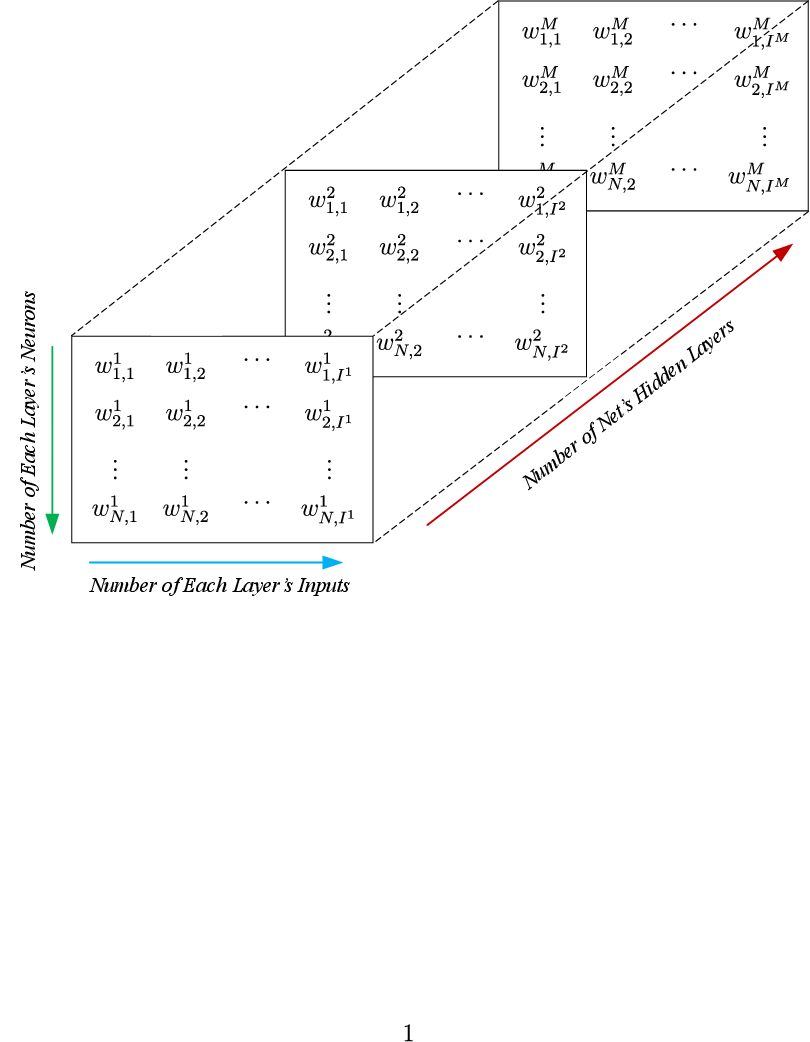} 
	\caption{Third-order tensor (3D matrix) of net's weights.}
	\label{fig:1}
\end{figure}

Generally, the behavior of a neural network is orchestrated by the behavioral resultant of its all neurons in different layers. From deep learning perspective, an output is obtained based upon the parameters and hyper-parameters of a model. The learning process in DNNs which implies the convergence to the exact solution within the grammar of numerical methods, is generally comprised of two algorithms, including forward pass and back-propagation algorithms \cite{ANN1}. The forward algorithm is thus arithmetically expressed for a multi-layer neural network with 'M' hidden layers as relation \eqref{eq:1}:

\begin{equation}
	\mathbf{u}^{m + 1} = \pmb{\sigma}^{m + 1}\left( \mathbf{W}^{m + 1}\mathbf{u}^{m}\mathbf{+}\mathbf{b}^{m + 1} \right)\mathbf{,\ \ \ \ \ \ \ \ \ \ }for\mathbf{\ }m = 0,\ 1,\ 2,\ 3,\ \ldots\ ,\ M - 1\mathbf{\ }
	\label{eq:1}
\end{equation}

\noindent where \textbf{W} is a third-order tensor (3D matrix) as depicted in Fig.~\ref{fig:1}, \textbf{b} and \pmb{$\sigma$} are the matrices of bias and activation function, and \textbf{X} and \textbf{u} are the vectors of input and output respectively. Moreover, the external inputs and outputs of the
DNN are generally defined as relations $\mathbf{u}^{\mathbf{0}}\mathbf{= X}$, and $\mathbf{u}_{\mathbf{net}}\mathbf{=}\mathbf{u}^{\mathbf{M}}$.

\begin{equation}
	\mathbf{X =}\begin{bmatrix}
		x_{1} \\
		x_{2} \\
		\vdots \\
		x_{I^1} \\
	\end{bmatrix}\mathbf{, \ \ \ \ \ \ \ \ \ b =}\begin{bmatrix}
		b_{1}^{1} & b_{1}^{2} & \ldots & b_{1}^{M} \\
		b_{2}^{1} & b_{2}^{2} & \ldots & b_{2}^{M} \\
		\vdots & \vdots & & \vdots \\
		b_{N}^{1} & b_{N}^{2} & \ldots & b_{N}^{M} \\
	\end{bmatrix} ,
	\label{eq:3}
\end{equation}
\\
\begin{equation}
	\mathbf{\pmb{\sigma} =}\begin{bmatrix}
		\sigma_{1}^{1} & \sigma_{1}^{2} & \ldots & \sigma_{1}^{M} \\
		\sigma_{2}^{1} & \sigma_{2}^{2} & \ldots & \sigma_{2}^{M} \\
		\vdots & \vdots & & \vdots \\
		\sigma_{N}^{1} & \sigma_{N}^{2} & \ldots & \sigma_{N}^{M} \\
	\end{bmatrix}\mathbf{,\ \ \ \ \ \ \ \ \ \ \ u =}\begin{bmatrix}
		u_{1}^{1} & u_{1}^{2} & \ldots & u_{1}^{M} \\
		u_{2}^{1} & u_{2}^{2} & \ldots & u_{2}^{M} \\
		\vdots & \vdots & & \vdots \\
		u_{N}^{1} & u_{N}^{2} & \ldots & u_{N}^{M} \\
	\end{bmatrix}
	\label{eq:4}
\end{equation}

\begin{figure}[H]
	\centering
	\subfloat {\includegraphics[width=1.25\textwidth]{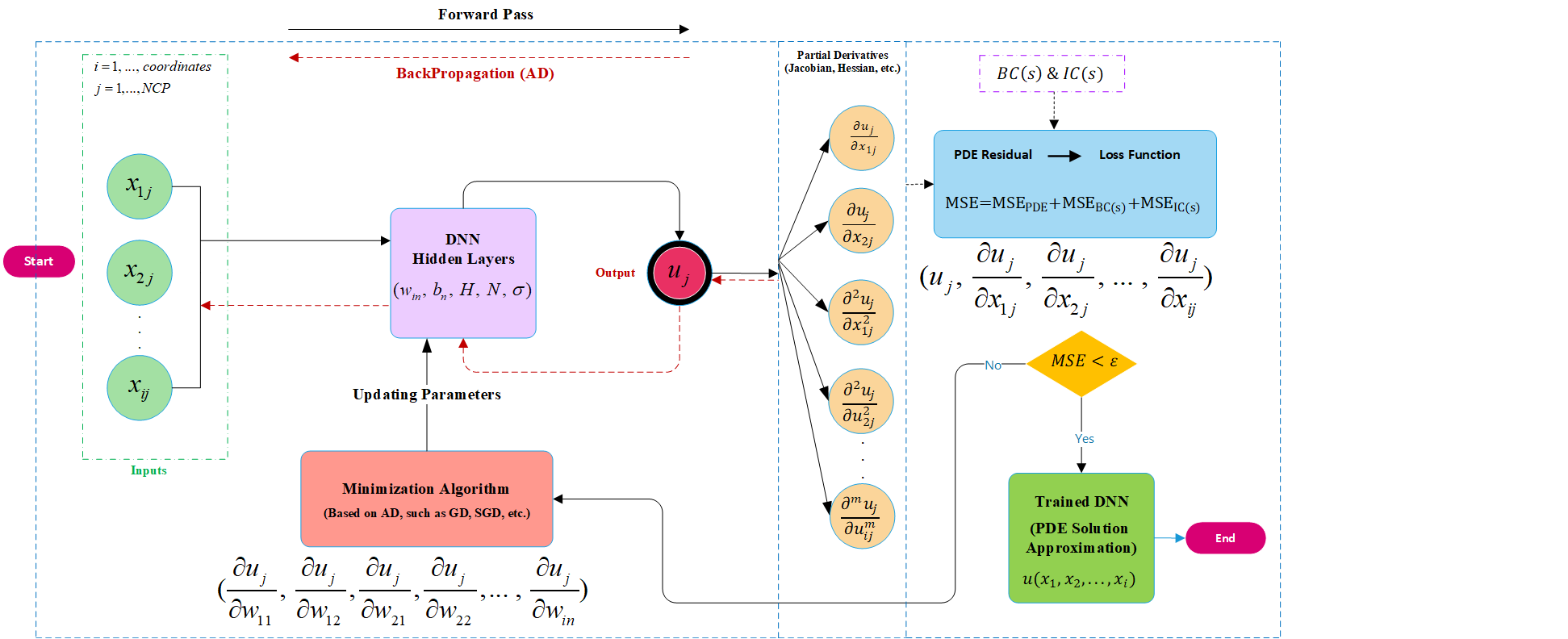}}
	\caption{Schematic of data-free physics-informed deep learning algorithm.}
	\label{fig:2}
\end{figure}

\noindent During backpropagation algorithm stage, which is the core of the training process, the model parameters are updated minimizing the loss function based upon the error associated with the previous forward step. In physics-informed deep learning, the loss function is modified to incorporate a physics-based term that penalizes residuals of the governing PDEs, boundary conditions (BCs), and/or initial conditions (ICs) \cite{PINN1}. The PDE residual represents the difference between its left-hand side (LHS) and right-hand side (RHS). Fig.~\ref{fig:2} meticulously illustrates the schematic representation of the algorithm used in data-free PINN (DF-PINN).

\noindent The total loss ($\mathcal{L}$) in a DF-PINN consists solely of terms enforcing physical laws and BCs/ICs \cite{raisi}. Without explicit data, the primary loss components are the PDE residual, $\mathcal{R}_{\text{PDE}}(\mathbf{x})$, and the BC/IC residuals, $\mathcal{R}_{\text{bc/ic}}(\mathbf{x})$, as defined in equation \eqref{eq:18}.

\begin{equation}
	\mathcal{L}_{\text{physics}} = \lambda_{\text{PDE}} \frac{1}{N_{\text{cp}}} \sum_{i=1}^{N_{\text{cp}}} \|\mathcal{R}_{\text{PDE}}(\mathbf{x}_i)\|^2 + \lambda_{\text{bc/ic}} \frac{1}{N_{\text{bc/ic}}} \sum_{i=1}^{N_{\text{bc/ic}}} \|\mathcal{R}_{\text{bc/ic}}(\mathbf{x}_i)\|^2
	\label{eq:18}
\end{equation}

\noindent where $\lambda$ is a hyperparameter that controls the importance of enforcing BCs/ICs relative to the PDE residual \cite{datafree}. 

\subsection{Physics-Informed Kolmogorov-Arnold Network (PIKAN)}

Physics-Informed Kolmogorov-Arnold Networks (PIKANs) have recently demonstrated their efficiency as a new PDE-solving paradigm by leveraging the Kolmogorov-Arnold representation theorem (KART). The Kolmogorov-Arnold representation theorem states that any multivariate function \( f: \mathbb{R}^n \to \mathbb{R} \) can be written as:

\begin{equation}
	f(\textbf{X}) = f(x_1, x_2, \dots, x_n) = \sum_{j=1}^{m} \psi_j \left( \sum_{i=1}^{n} \phi_{ij} (x_i) \right)
\end{equation}

\noindent where \( \psi_i \) and \( \phi_{ij} \) are continuous univariate functions. This theorem suggests that high-dimensional function learning can be efficiently handled by decomposing the target function into a hierarchy of univariate transformations. For k-th layer in a Kolmogorov-Arnold Network (KAN), assuming an identity mapping for the function $\psi$ such that $ \psi(z)=z$, the $\phi$ function can be rewritten in matrix form as:

\begin{equation}
	\Phi^{k} =
	\begin{bmatrix}
		\phi_{11} & \phi_{12} & \cdots & \phi_{1n} \\
		\phi_{21} & \phi_{22} & \cdots & \phi_{2n} \\
		\vdots & \vdots & \ddots & \vdots \\
		\phi_{m1} & \phi_{m2} & \cdots & \phi_{mn}
	\end{bmatrix}
\end{equation}

\noindent Therefore, a full single-output KAN network with \( L \) layers is expressed as:

\begin{equation}
	f(x_1, x_2, \dots, x_n) =
	\sum_{i_{L}=1}^{n_{L}} \phi^{L}_{i_{L+1},i_{L}}
	\left( \sum_{i_{L-1}=1}^{n_{L-1}} \dots 
	\left( \sum_{i_2=1}^{n_2} \phi^{2}_{2,i_3,i_2}
	\left( \sum_{i_1=1}^{n_1} \phi^{1}_{i_2,i_1}(x_{i_1}) \right) \right) \right)
\end{equation}

\noindent In conventional neural networks such as multi-layer perceptrons (MLPs), activation functions are uniformly applied across all nodes within the network. However, activation functions in KANs are associated with the edges rather than the nodes. These edges, which correspond to the weight connections in traditional neural networks, each possess distinct activation functions, allowing for a more adaptive function representation. These activation functions can be represented as a weighted combination of basis functions and B-splines, which are defined as:

\begin{equation}
	\phi(x) = w_b b(x) + w_s \sum_{i} c_i B_i(x),
\end{equation}

\noindent where \( b(x) \) is a predefined basis function such as Sigmoid Linear Unit (SiLU), \( B_i(x) \) are B-spline basis functions, and \( w_b, w_s, c_i \) are learnable parameters. The SiLU, also known as the Swish activation function, is defined as:

\begin{equation}
	\text{SiLU}(x) = \frac{x}{1 + e^{-x}}
\end{equation}

\begin{figure}[H]
	\centering
	\includegraphics[width=0.99\textwidth]{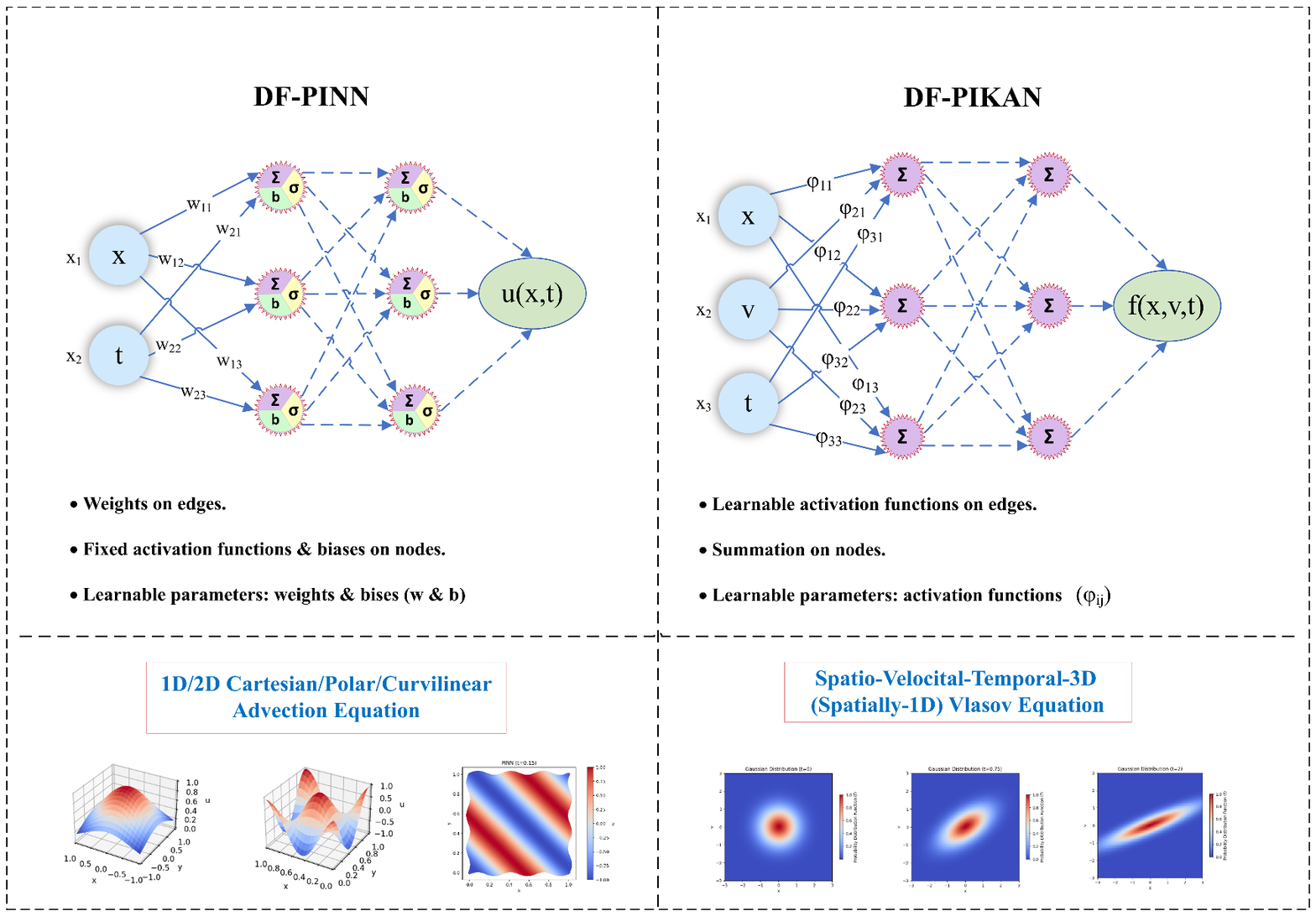} 
	\caption{Schematic of space-time DF-PINN and phase-space DF-PIKAN architectures.}
	\label{fig:3}
\end{figure}

\noindent In this study, we extend the framework of PINNs by employing PIKANs to efficiently approximate the phase-space solutions of Collisionless Boltzmann Equation (CBE) in a continuous-velocity manner. This approach enables a decomposition of the high-dimensional solution space into simpler, learnable univariate transformations, leading to enhanced interpretability in approximating high-dimensional PDE solutions. Fig.~\ref{fig:3} presents a comparative analysis of DF-PINN in space-time and continuous-velocity data-free PIKAN (DF-PIKAN) in phase-space.

\subsection{Feature Embeddings via HWF-PIKAN}

Although PIKAN architectures provide improved functional decomposition compared to standard PINNs, both models may still exhibit spectral bias, learning low-frequency components more easily than highly oscillatory or sharply localized features. To mitigate this limitation, the inputs are first mapped into a higher-dimensional spectral-multi-resolution feature space before being passed to the network. This preprocessing enhances the model's ability to represent complex solution structures without increasing the depth or width of the network.

\noindent Let the normalized scalar input be $u \in [0,1]$, corresponding to one coordinate of the physical variable (e.g., $y$ or $v$). The proposed Hybrid Wavelet-Fourier (HWF) embedding constructs a multi-resolution representation by combining global Fourier features with localized Ricker wavelets.

\paragraph{Fourier embedding.}  
For each scalar input $u$, the Fourier component of the embedding is defined as
\begin{equation}
	\gamma_F(u)
	=
	\left[
	\sin(2\pi k u),\;
	\cos(2\pi k u)
	\right]_{k=1}^{M_{\mathrm{Fourier}}},
\end{equation}
where $M_{\mathrm{Fourier}}$ is the number of sinusoidal frequency pairs.  
Unlike random Fourier features, the implementation uses deterministic integer frequencies, providing smooth global variations across the domain.

\paragraph{Wavelet embedding.}  
To incorporate localized and multi-scale structure, a set of Ricker (Mexican-hat) wavelets is applied:
\begin{equation}
	\psi_{j,k}(u)
	=
	\left(1 - \left(\frac{u - c_{j,k}}{s_j}\right)^2 \right)
	\exp\!\left(
	-\frac{1}{2}\left(\frac{u - c_{j,k}}{s_j}\right)^2
	\right),
\end{equation}
where $j=0,\dots,J_{\mathrm{wavelet}}-1$ indexes resolution scales,  
$c_{j,k}$ are uniformly spaced centers at scale $j$,  
and $s_j = \mathcal{O}(2^{-j})$ is a decreasing scale parameter that controls spatial localization.  
This construction yields $2^j$ wavelets per scale, providing a total of $2^{J_{\mathrm{wavelet}}}-1$ localized features.

\paragraph{Hybrid embedding.}  
The full one-dimensional HWF embedding is obtained by concatenating the Fourier and wavelet components:
\begin{equation}
	\gamma_{\mathrm{HWF}}(u)
	=
	\big[
	\gamma_F(u),\;
	\gamma_W(u)
	\big].
\end{equation}
For multi-dimensional PDEs, the embedding is applied independently to each coordinate (e.g., $y$ and $v$), and the resulting feature vectors are concatenated before being fed into the KAN/B-spline core.

\begin{figure}[H]
	\centering
	\includegraphics[width=1\textwidth]{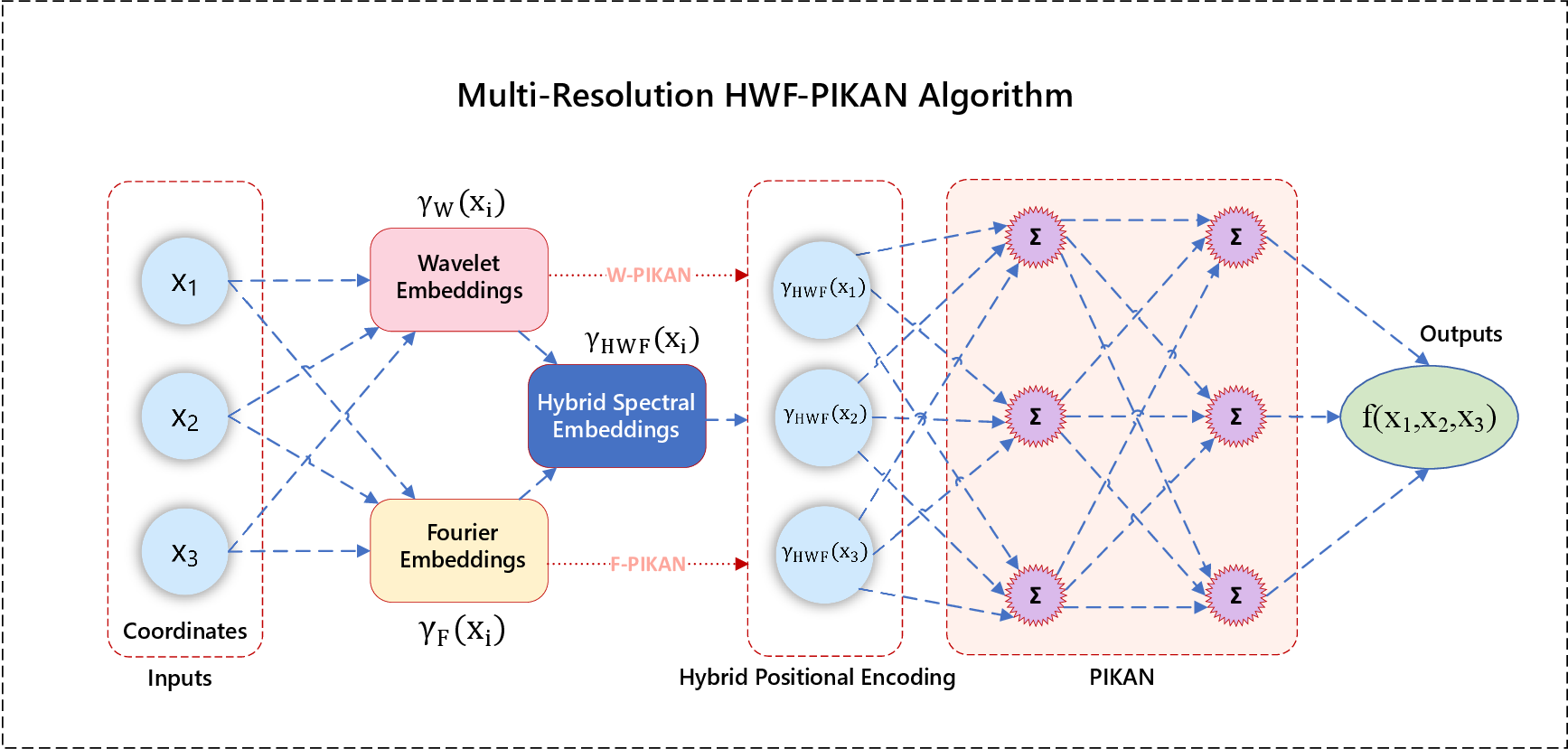} 
	\caption{Schematic of the proposed multi-resolution HWF-PIKAN architecture.}
	\label{fig:4}
\end{figure}

\subsection{HWF-PIKAN Implementation}
\label{sec:software}
This hybrid representation allows the network to efficiently capture both globally smooth structures and sharp, localized variations, making it particularly effective for advection-dominated and multi-scale PDEs. The following pseudo-code outlines the general workflow of the HWF–PIKAN framework.

\begin{algorithm}[H]
	\caption{General HWF-PIKAN Algorithm for Solving PDEs}
	\label{alg:hwf-pikan-general}
	\begin{algorithmic}[1]
		\Require PDE operator \(\mathcal{N}[u](\mathbf{z})\), IC/BC \(u_0, g\), domain \(\Omega\)
		\Require Collocation sets: interior \(\mathcal{Z}_r\), initial \(\mathcal{Z}_{\mathrm{IC}}\), boundary \(\mathcal{Z}_{\mathrm{BC}}\)
		\Require Hyperparameters: learning rate \(\eta\), HWF and KAN settings, loss weights \(\lambda_{\mathrm{PDE}}, \lambda_{\mathrm{IC}}, \lambda_{\mathrm{BC}}\)
		
		\State Initialize HWF--PIKAN parameters \(\theta\)
		
		\Statex \Comment{HWF--PIKAN mapping}
		\ForAll{\(\mathbf{z} \in \Omega\)}
		\State Normalize \(\mathbf{z} \mapsto \tilde{\mathbf{z}}\)
		\State Compute Fourier features \(\gamma_F(\tilde{\mathbf{z}})\) and wavelet features \(\gamma_W(\tilde{\mathbf{z}})\)
		\State Form hybrid embedding \(\gamma_{\mathrm{HWF}} = [\gamma_F, \gamma_W]\)
		\State \(\hat{u}_\theta(\mathbf{z}) \gets \mathrm{KAN\_Core}(\mathrm{LayerNorm}(\gamma_{\mathrm{HWF}}))\)
		\EndFor
		
		\Statex \Comment{Losses}
		\State \(\mathcal{L}_{\mathrm{PDE}} = \mathrm{MSE}(\mathcal{N}[\hat{u}_\theta](\mathbf{z}), 0), \; \mathbf{z} \in \mathcal{Z}_r\)
		\State \(\mathcal{L}_{\mathrm{IC}}  = \mathrm{MSE}(\hat{u}_\theta(\mathbf{x},0), u_0(\mathbf{x})), \; \mathbf{x} \in \mathcal{Z}_{\mathrm{IC}}\)
		\State \(\mathcal{L}_{\mathrm{BC}}  = \mathrm{MSE}(\hat{u}_\theta(\mathbf{x},t), g(\mathbf{x},t)), \; (\mathbf{x},t) \in \mathcal{Z}_{\mathrm{BC}}\)
		\State \(\mathcal{L} = \lambda_{\mathrm{PDE}}\mathcal{L}_{\mathrm{PDE}} + \lambda_{\mathrm{IC}}\mathcal{L}_{\mathrm{IC}} + \lambda_{\mathrm{BC}}\mathcal{L}_{\mathrm{BC}}\)
		
		\Statex \Comment{Training: IC warm-up, then full PIKAN + optional L-BFGS}
		\For{IC iterations}
		\State Sample \(\mathcal{B}_{\mathrm{IC}} \subset \mathcal{Z}_{\mathrm{IC}}\), update \(\theta\) with Adam on \(\mathcal{L}_{\mathrm{IC}}\)
		\EndFor
		\For{Adam iterations}
		\State Sample \(\mathcal{B}_r, \mathcal{B}_{\mathrm{IC}}, \mathcal{B}_{\mathrm{BC}}\), update \(\theta\) with Adam on \(\mathcal{L}\)
		\EndFor
		\For{L-BFGS epochs (optional)}
		\State Use closure on (full or large) batches from \(\mathcal{Z}_r, \mathcal{Z}_{\mathrm{IC}}, \mathcal{Z}_{\mathrm{BC}}\) and take one L-BFGS step
		\EndFor
		
		\State \Return trained HWF-PIKAN model \(\hat{u}_\theta\)
	\end{algorithmic}
\end{algorithm}

\noindent All numerical experiments reported in this work were carried out using a standalone open-source Python implementation of the proposed Hybrid Wavelet-Fourier-Enhanced Physics-Informed Kolmogorov--Arnold Network (HWF-PIKAN) framework. The implementation is written in \texttt{Python~3.x} and makes use of \texttt{PyTorch} for automatic differentiation and GPU-accelerated training, together with \texttt{NumPy} and \texttt{Matplotlib} for data handling and visualization. The codebase is modular and allows straightforward adaptation to a broad range of partial differential equations (PDEs), geometry configurations, and boundary conditions.
The complete prototype implementation of the HWF-PIKAN framework, including example
scripts for the 1D advection equation, 2D advection equation, and the collisionless Boltzmann
equation (CBE), is publicly available at:
\textbf{\url{https://github.com/m-heravifard/HWF-PIKAN}}.

\subsubsection{Network Architecture}

The implementation adopts the general HWF-PIKAN structure described earlier and consists of three components. 

\noindent \textbf{(i) Hybrid spectral embedding (HWF):}
Each input coordinate is normalized to \([0,1]\) and mapped into a hybrid 
Fourier-wavelet feature space. For a scalar input \(u\), the embedding is as follows:

\begin{equation}
\gamma_{\mathrm{HWF}}(u)
= \big[\sin(2\pi k u),\, \cos(2\pi k u)\big]_{k=1}^{M_{\mathrm{Fourier}}}
\ \cup\ 
\{\psi_{j,k}(u)\}_{j=0}^{J_{\mathrm{wavelet}}-1},
\end{equation}

\noindent where \(\psi_{j,k}\) are multiresolution Ricker wavelets centered at uniformly spaced 
locations. For multi-dimensional inputs, embeddings of all coordinates are concatenated.

\noindent \textbf{(ii) KAN / B-spline core:}
The hybrid features are then passed through a Kolmogorov--Arnold network with a 
B-spline nonlinearity. A linear mapping first produces \(Q\) channels, which are squashed 
into \((0,1)\) using a scaled \(\tanh\). Each channel is expanded over cubic B-spline basis 
functions \(B_i(s)\) defined on an open-uniform knot vector with \(n_{\mathrm{control}}\) control 
points. Learnable spline coefficients and a final linear combination yield the network output 
\(u_\theta(\mathbf{z})\).

\noindent \textbf{(iii) Normalization and modular composition:}
Layer normalization is applied to the hybrid embeddings, and the framework allows optional 
input preprocessing such as coordinate transforms or dimension reduction, depending on the 
PDE under consideration.

Typical hyperparameters are:
\[
M_{\mathrm{Fourier}} = 32,\quad
J_{\mathrm{wavelet}} = 3,\quad
Q = 16,\quad
p = 3,\quad
n_{\mathrm{control}} = 25.
\]

\subsubsection{Loss function and optimization schedule}

The physics-informed loss is defined as

\begin{equation}
\mathcal{L}
= \lambda_{\mathrm{PDE}}\,\mathcal{L}_{\mathrm{PDE}}
+ \lambda_{\mathrm{IC}}\,\mathcal{L}_{\mathrm{IC}}
+ \lambda_{\mathrm{BC}}\,\mathcal{L}_{\mathrm{BC}},
\end{equation}

\noindent where
\[
\mathcal{L}_{\mathrm{PDE}} = \mathrm{MSE}(\mathcal{N}[u_\theta](\mathbf{z}),0),\quad
\mathcal{L}_{\mathrm{IC}} = \mathrm{MSE}(u_\theta(\mathbf{x},0),u_0(\mathbf{x})),\quad
\mathcal{L}_{\mathrm{BC}} = \mathrm{MSE}(u_\theta(\mathbf{x},t),g(\mathbf{x},t)).
\]

\noindent Training proceeds in three stages to improve convergence:  
(1) an IC warm-up phase minimizing \(\mathcal{L}_{\mathrm{IC}}\);  
(2) full training with the Adam optimizer on the complete loss; and  
(3) an optional L--BFGS refinement to further reduce residuals.  
This schedule consistently stabilizes training and accelerates convergence.

\section{Numerical Results}
	In this section, the achieved results based on the physics-driven deep learning methods for several cases are presented. It should be noted that no previous data has been used in the implemented solvers except for the boundary/initial conditions.
	
\subsection{Advection Problems in Space-Time}
	First of all, in order to implement the HWF-PIKAN solver, the advection (transport) equation is chosen as the first case \cite{advection}. The primary focus of solving this widely-used PDE often revolves around discontinuous shock solutions, which pose significant challenges for conventional numerical schemes. The advection PDE governs the motion of a conserved quantity that is advected by a velocity vector field. This quantity can even represent a probability density function within the context of statistical physics, considering the Collisionless Boltzmann equation (CBE) in phase-space, which will be addressed in the next subsection.\\ 
	
\begin{figure}[H]
	\centering

	\subfloat [Case 1] 
	{\includegraphics[width=0.28\textwidth]{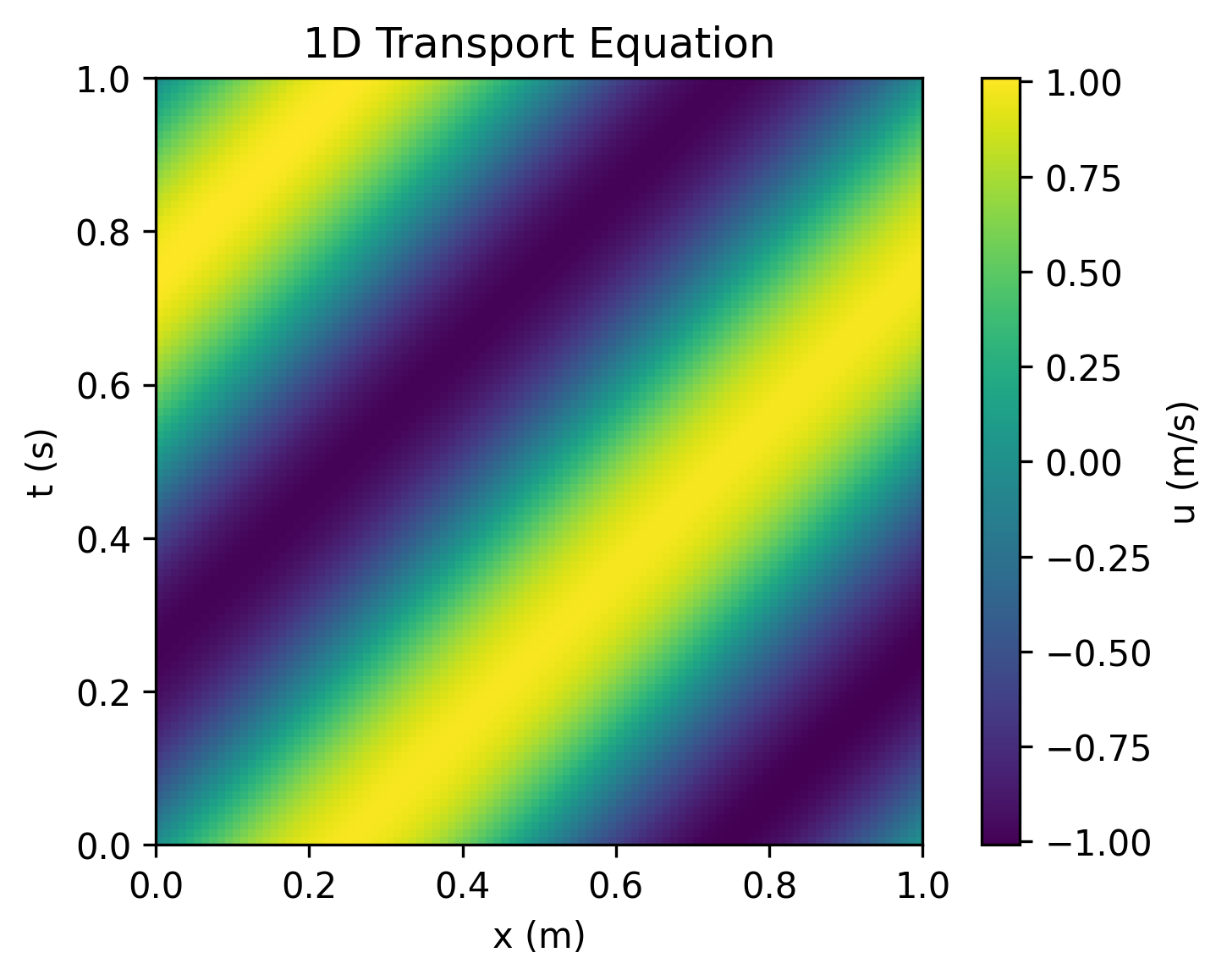}}
	\subfloat [Case 2] 
	{\includegraphics[width=0.28\textwidth]{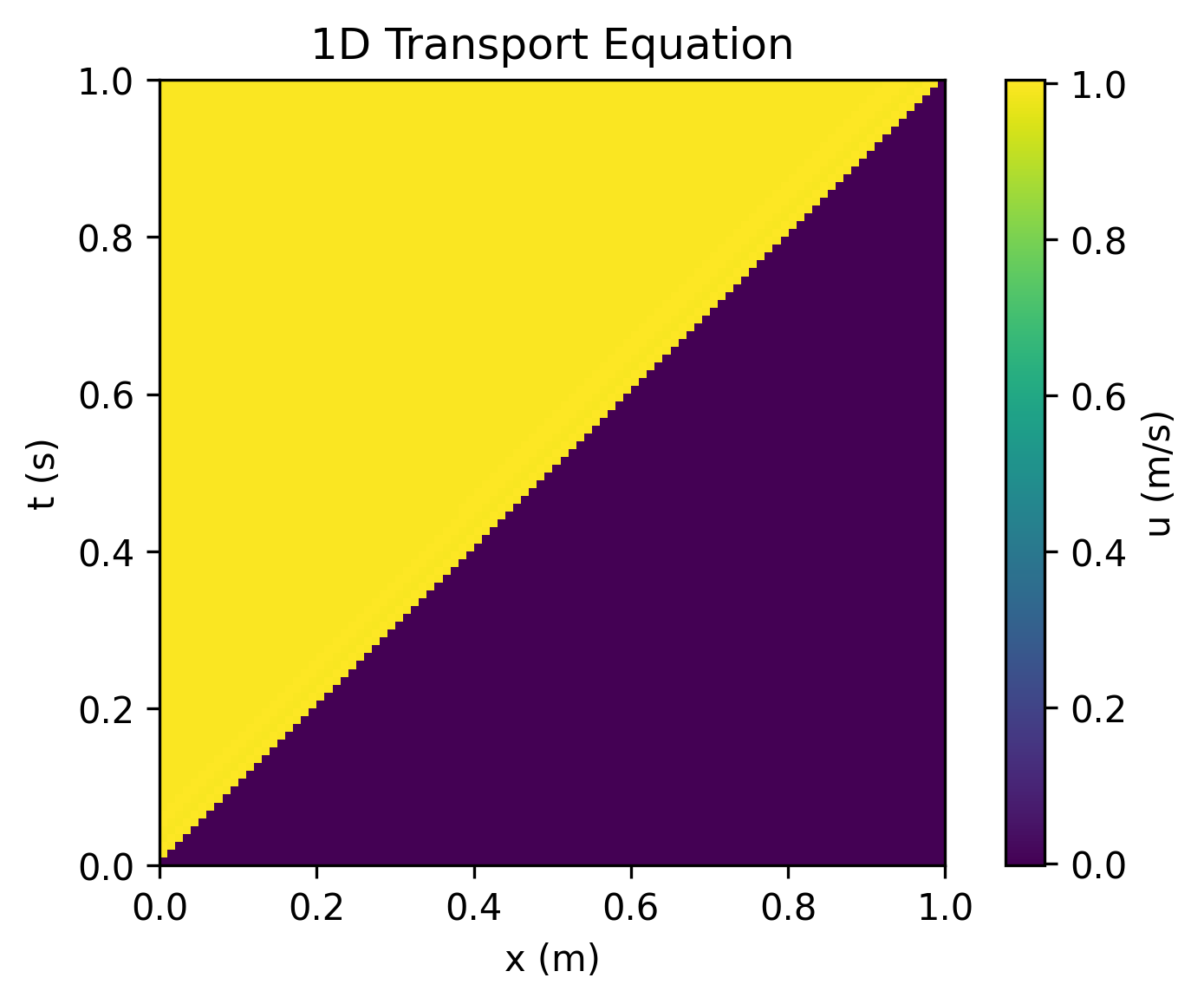}}
	\subfloat [Collocation points] 
	{\includegraphics[width=0.285\textwidth]{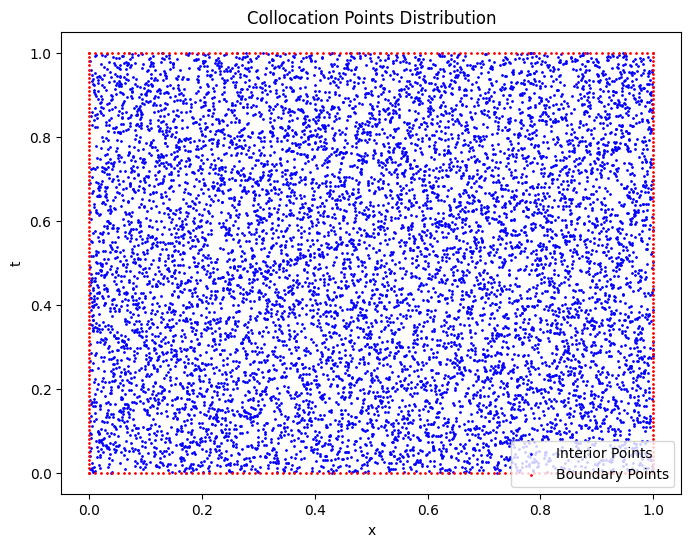}}
	\\ 

	\subfloat [Case 3 (t=0)]{\includegraphics[width=0.22\textwidth]{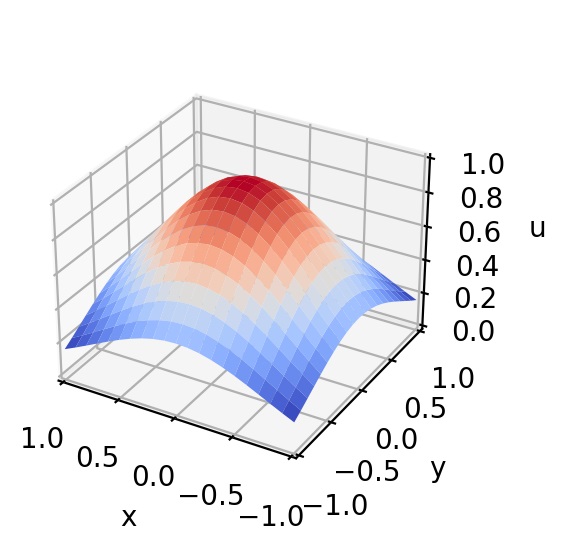}} 
	\subfloat [Case 3 (t=0.5)]{\includegraphics[width=0.22\textwidth]{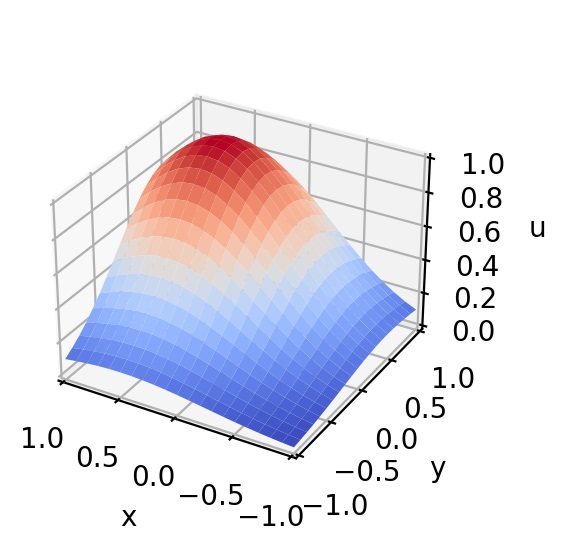}} 
	\subfloat [Case 4 (t=0)]{\includegraphics[width=0.22\textwidth]{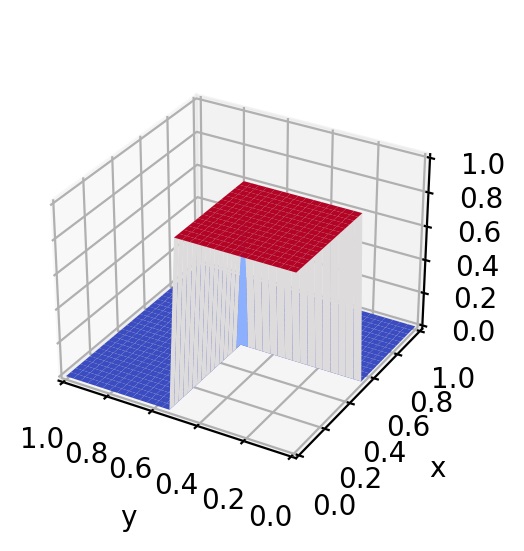}}
	\subfloat [Case 4 (t=0.2)]{\includegraphics[width=0.22\textwidth]{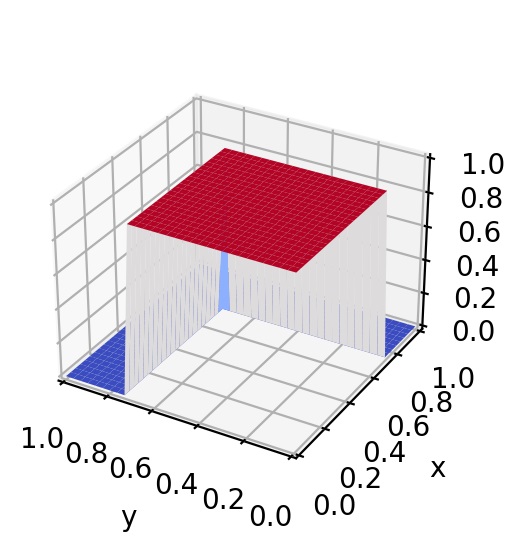}}
	\\
	\subfloat [Case 3 (t=0)]{\includegraphics[width=0.22\textwidth]{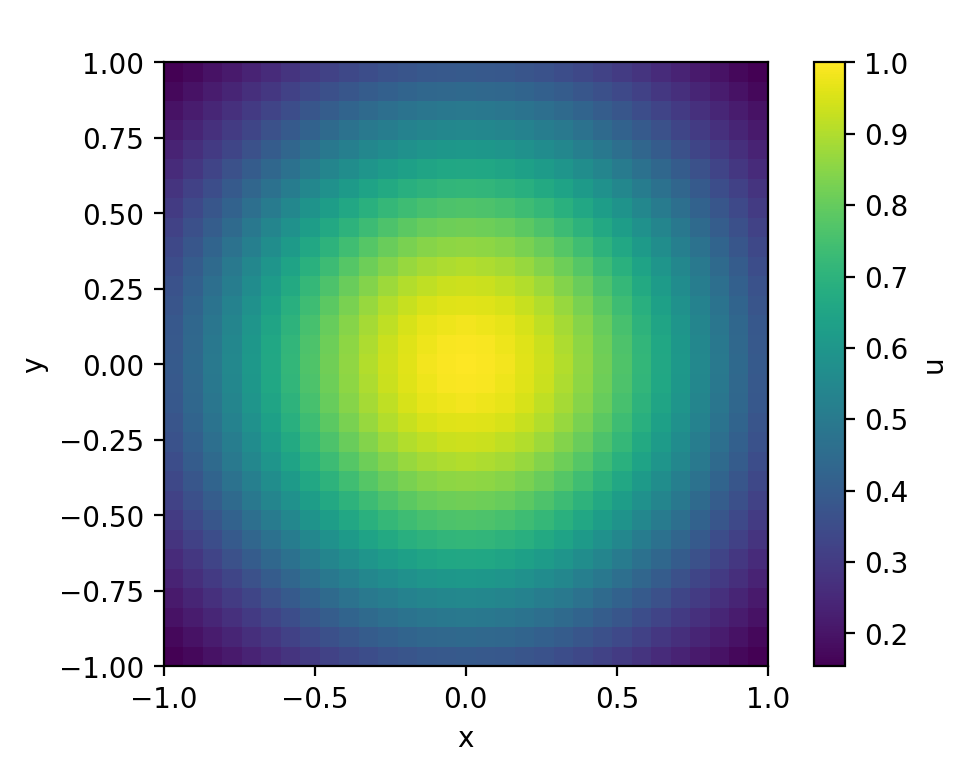}}
	\subfloat [Case 3 (t=0.5)]{\includegraphics[width=0.22\textwidth]{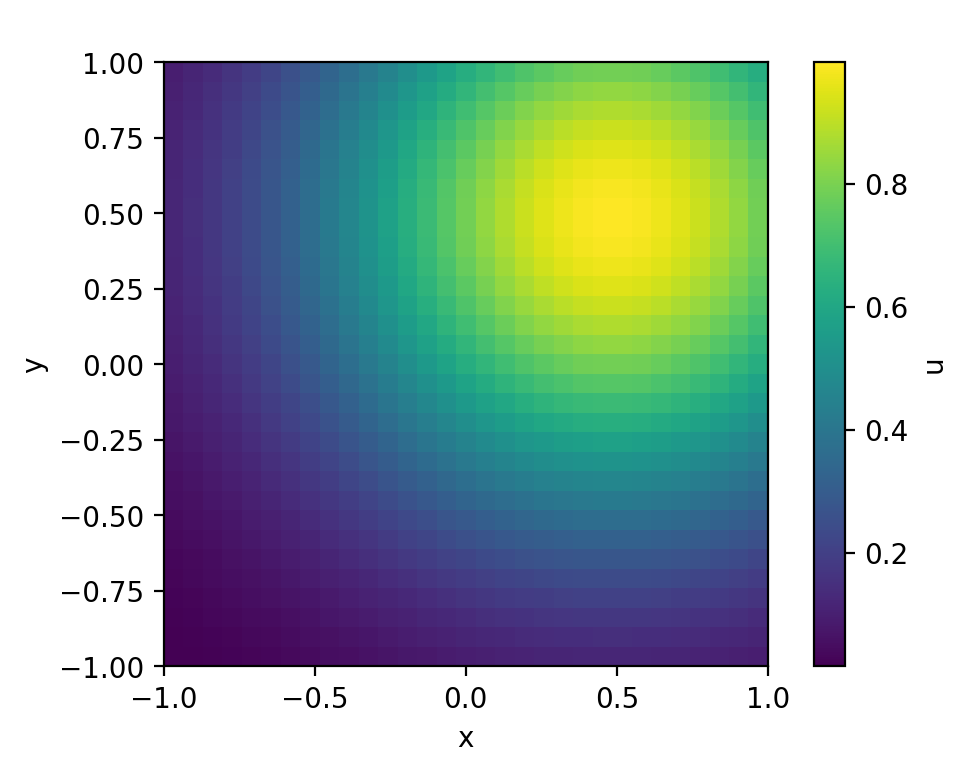}}
	\subfloat [Case 4 (t=0)]{\includegraphics[width=0.22\textwidth]{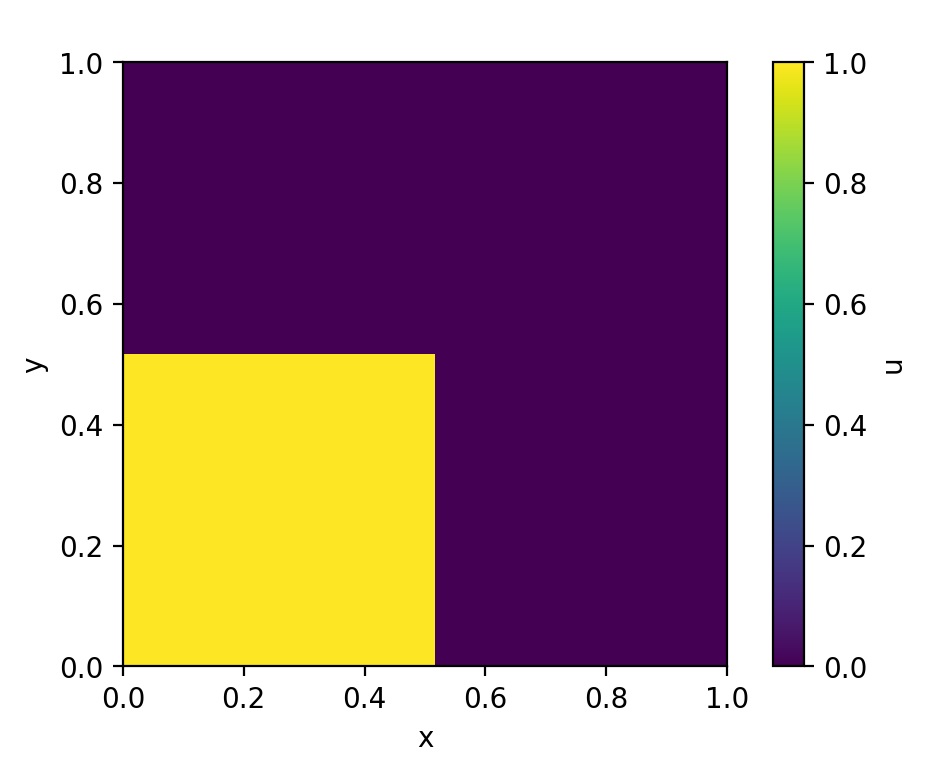}}
	\subfloat [Case 4 (t=0.2)]{\includegraphics[width=0.22\textwidth]{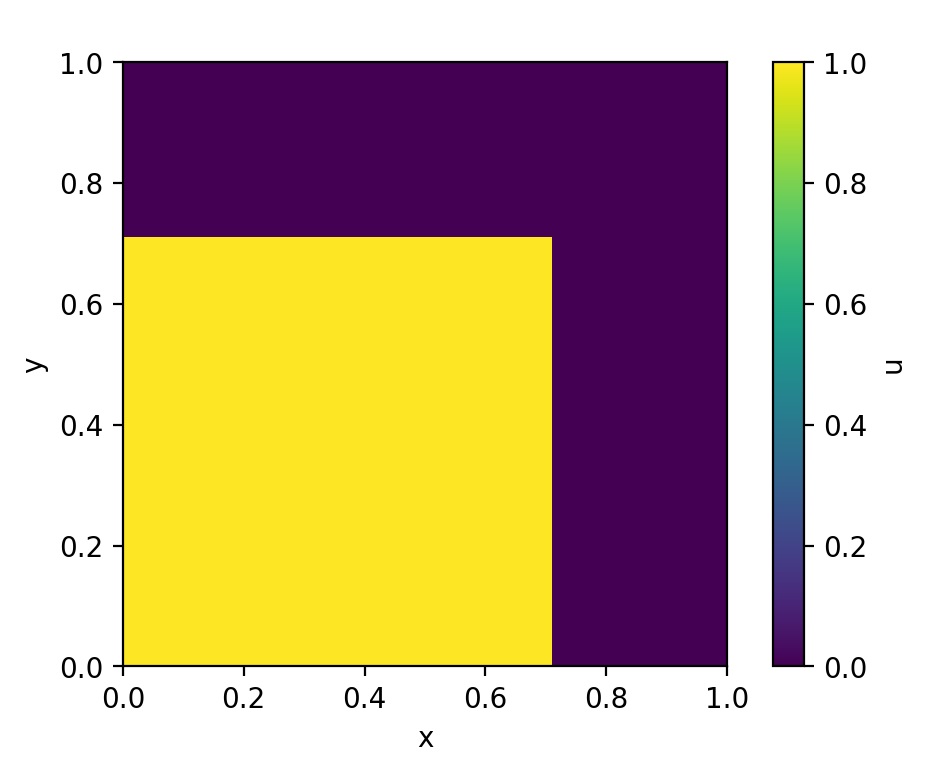}}
	\caption{HWF-PIKAN solutions for 1D/2D advection equations (Continuous and heaviside).}
	\label{fig:5}
\end{figure}

The problem of 1D advection equation with a sinusoidal initial disturbance which is an initial value problem (IVP) is defined as relation \eqref{eq:21} and the results are compared with exact solution which show appropriate agreement, as given in Figs.~\ref{fig:5} (a-b).

	\begin{equation}
	Case \,1:\ \left\{ \begin{matrix}
		u_{t} + cu_{x} = 0 \\
		u(x,0) = \sin(2\pi x) \\
	\end{matrix} \right.\ 
	\label{eq:21}
	\end{equation}

\noindent	Secondly, the results of the implemented solver considering a discontinuous initial perturbation for the advection equation is presented in Figs~\ref{fig:5} (c-d). In this case, a Heaviside (step) function, which has a discontinuity at $x=0.5$, is considered as the initial condition as equation \eqref{eq:22}.

\begin{equation}
	Case \,2:\ \left\{ \begin{matrix}
		u_{t} + cu_{x} = 0 \\
		u(x,0) = \left\{ \begin{matrix}
			1\ \ \ \ \ \ if\ x < 0.5 \\
			0\ \ \ \ \ \ if\ x \geq 0.5 \\
		\end{matrix} \right.\  \\
	\end{matrix} \right.\
	\label{eq:22} 
\end{equation}

\noindent	In Figs.~\ref{fig:5} (c-d), the first graph is related to the right-running wave (c=1) and the next graph is dedicated to the left-running movement (c=-1). According to the results obtained in this case, the HWF-PIKAN method can be considered as an appropriate method for resolving discontinuities in the computational domain. Moreover, the space-time contours in Figs.~\ref{fig:5} (e-f) indicate numerical solutions for all times at once. Fig.~\ref{fig:5} (g) also visualizes the random collocation points distribution.\\ 	

On the other hand, we present the results obtained from solving the 2D advection equation with both continuous and discontinuous initial conditions (see \cite{2Dadvection1, 2Dadvection2,2Dadvection3, 2Dadvection4}). As the continuous case here, the initial condition is represented by a smooth, bell-shaped Gaussian pulse, which is given by the expression \eqref{eq:23}. The results are shown in Figs.~\ref{fig:5} (h-i, l-m).

\begin{equation}
	Case \,3:\ \left\{ \begin{matrix}
		u_{t} + c(u_{x} + u_{y}) = 0 \\
		u(x,y,0) = exp( - x^{2} - y^{2}) \\
	\end{matrix} \right.\ 
	\label{eq:23}	
\end{equation}

%
%
%
%

\noindent In addition, a discontinuous initial condition in the form of a 2D Heaviside (step) function as written in relation \eqref{eq:26} is considered and the results are as Figs.~\ref{fig:5} (j-k, n-o).

\begin{equation}
	Case \,4:\ \left\{ \begin{matrix}
		u_{t} + c(u_{x} + u_{y}) = 0 \\
		u(x,y,0) = \left\{ \begin{matrix}
			1\ \ \ \ \ \ if\ (x < 0.5) \\
			0\ \ \ \ \ \ \ \ otherwise \\
		\end{matrix} \right.\  \\
	\end{matrix} \right.\ \ and\ \ (y < 0.5)
	\label{eq:26}	
\end{equation}

\begin{figure}[H]
	\centering
	\subfloat[t=0.25\label{fig:b}]{\includegraphics[width=0.22\textwidth]{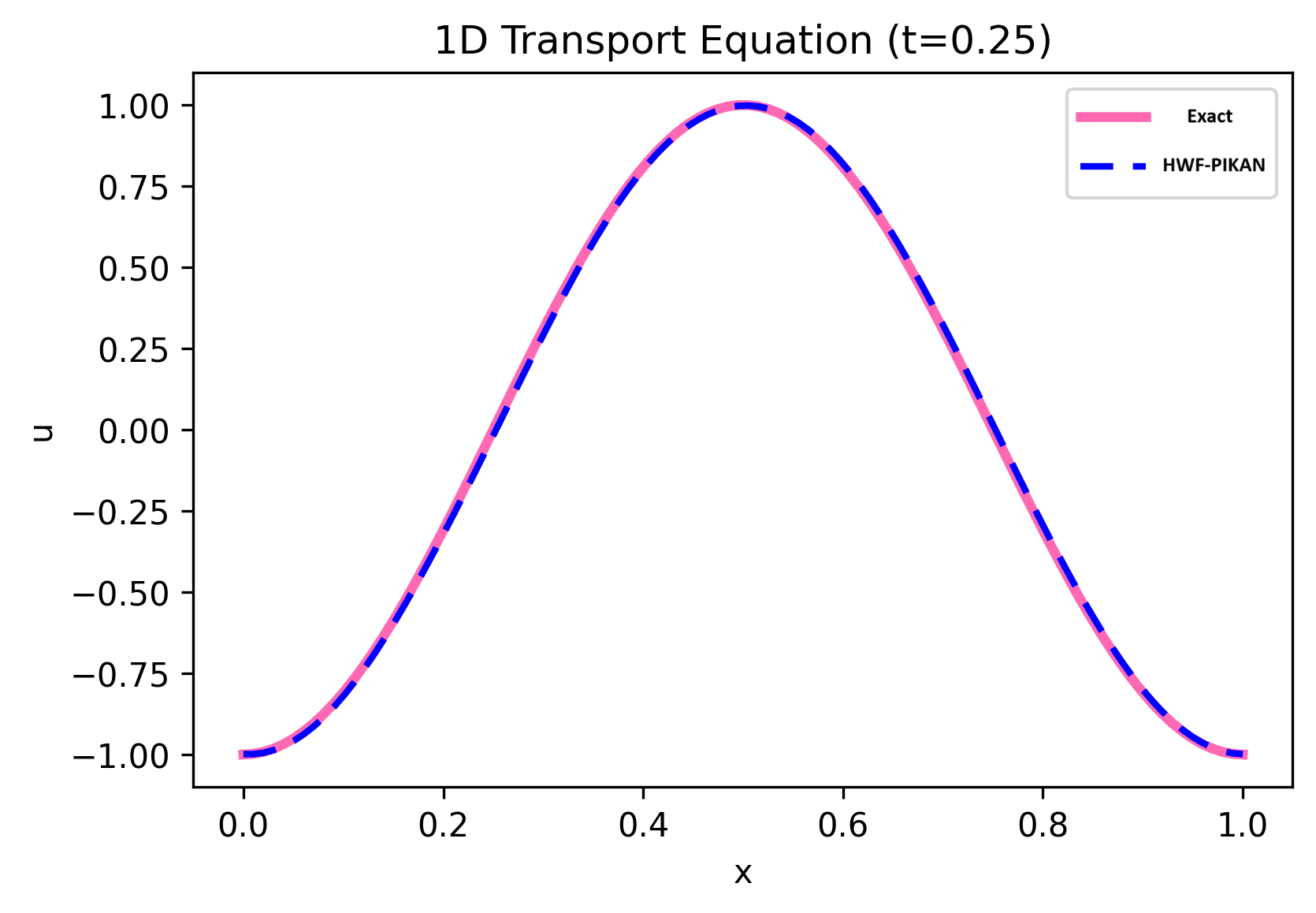}} 
	\subfloat[t=0.6\label{fig:c}]{\includegraphics[width=0.22\textwidth]{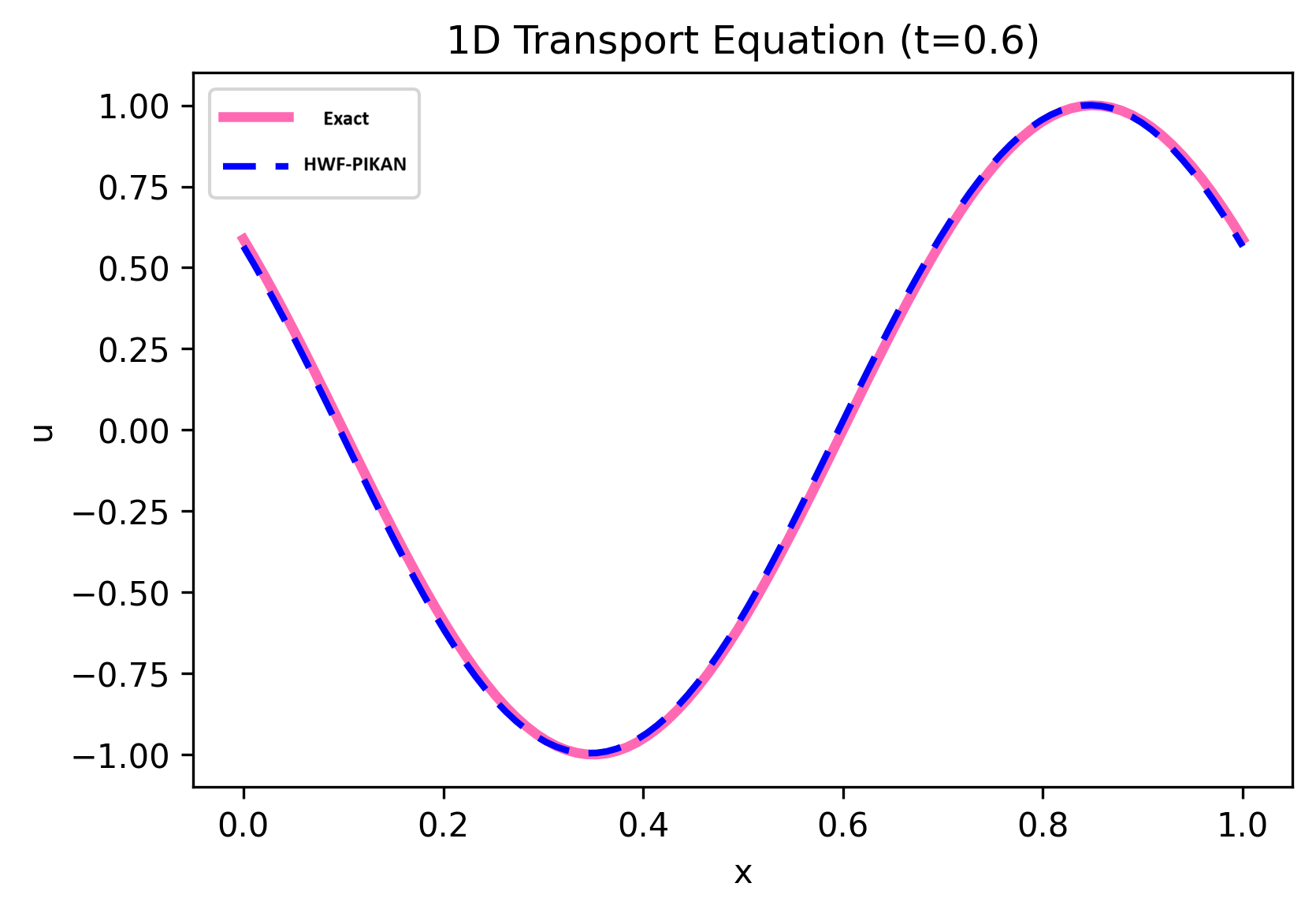}}
	\subfloat[t=0.2, c=1]{\includegraphics[width=0.22\textwidth]{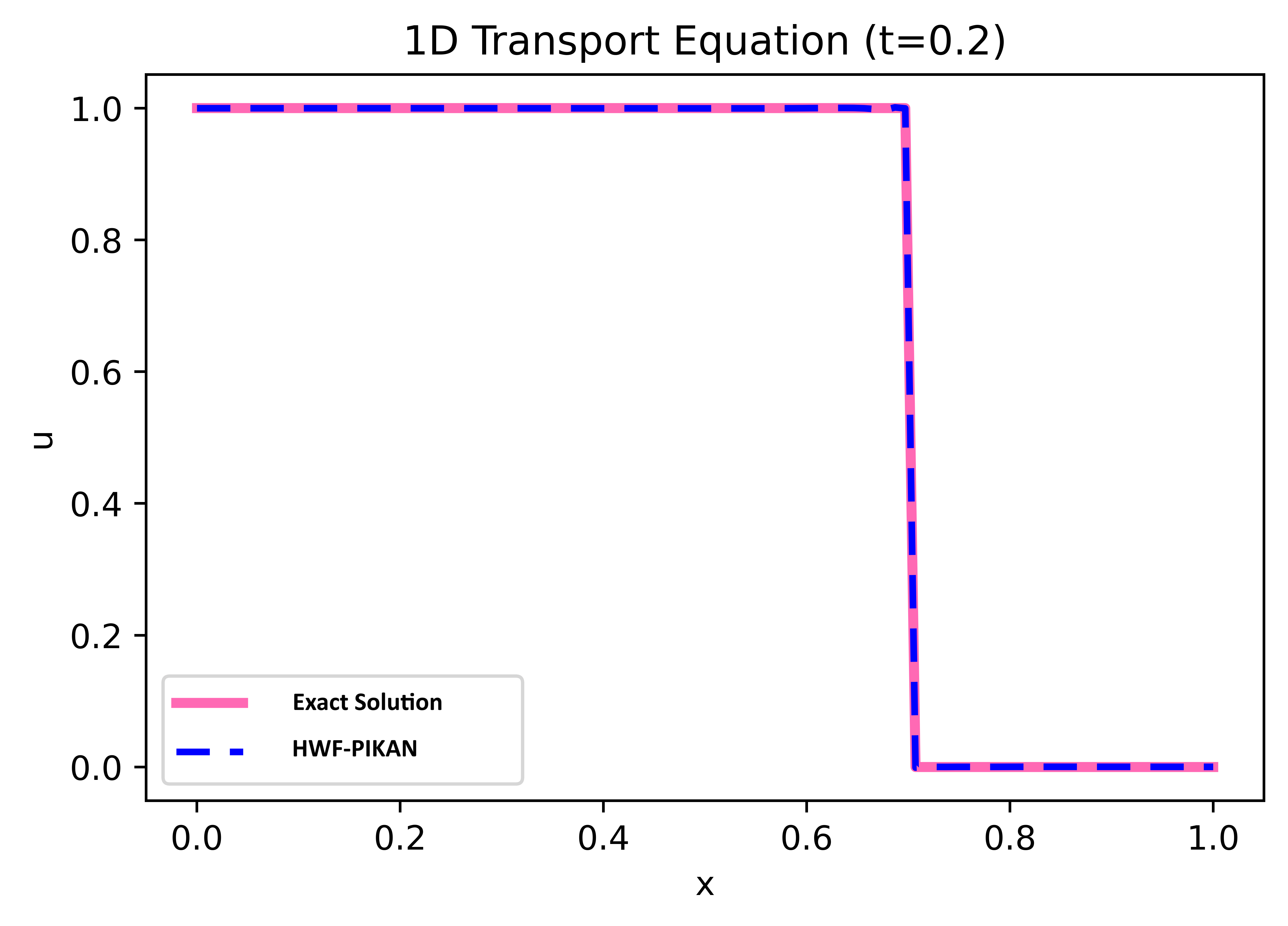}\label{fig:f}}
	\subfloat[t=0.3, c=-1]{\includegraphics[width=0.22\textwidth]{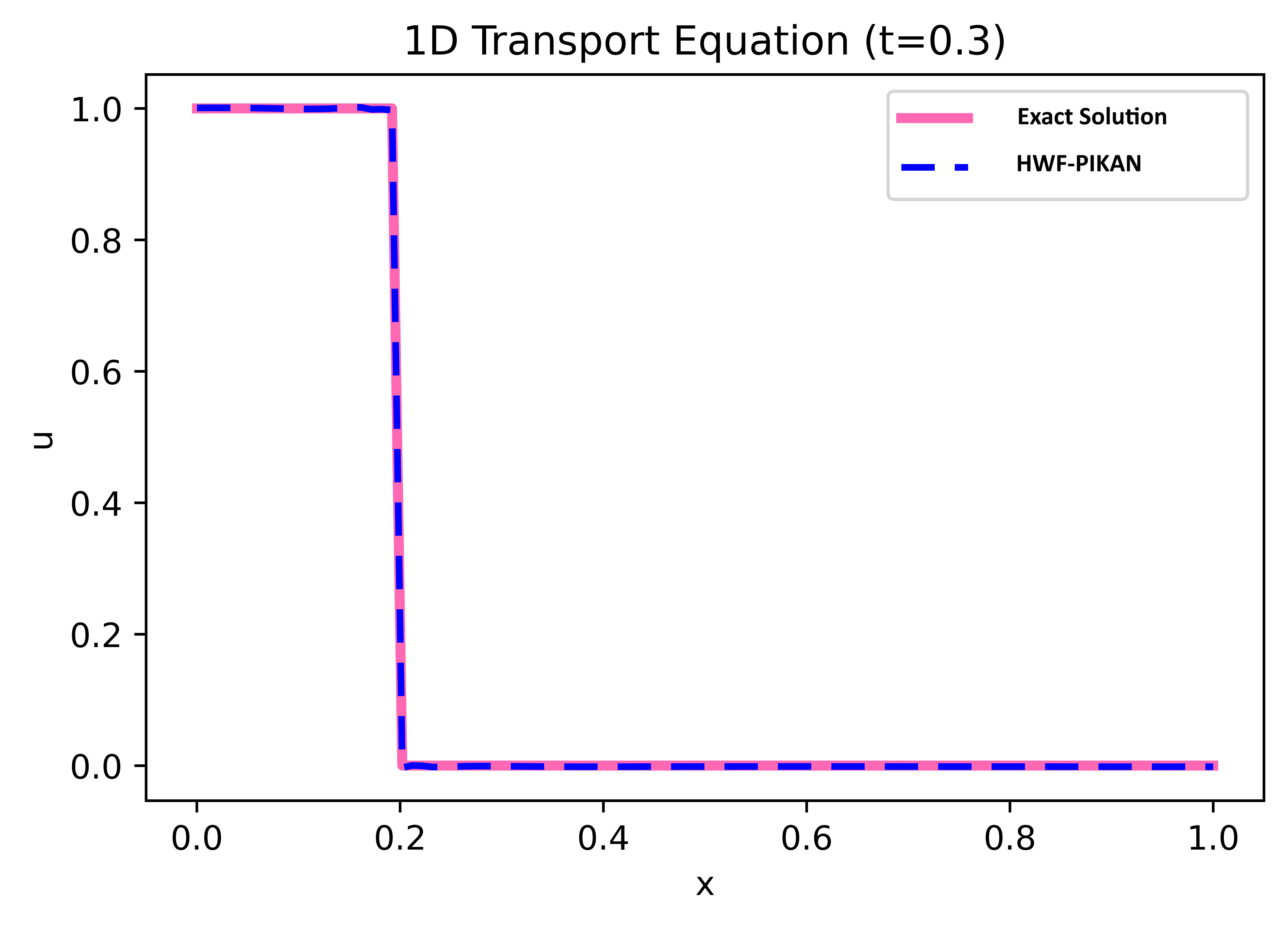}\label{fig:h}}
\end{figure}

\begin{figure}[H]
	\centering
	\subfloat [Ablation Study (Case 1)]{\includegraphics[width=0.46\textwidth]{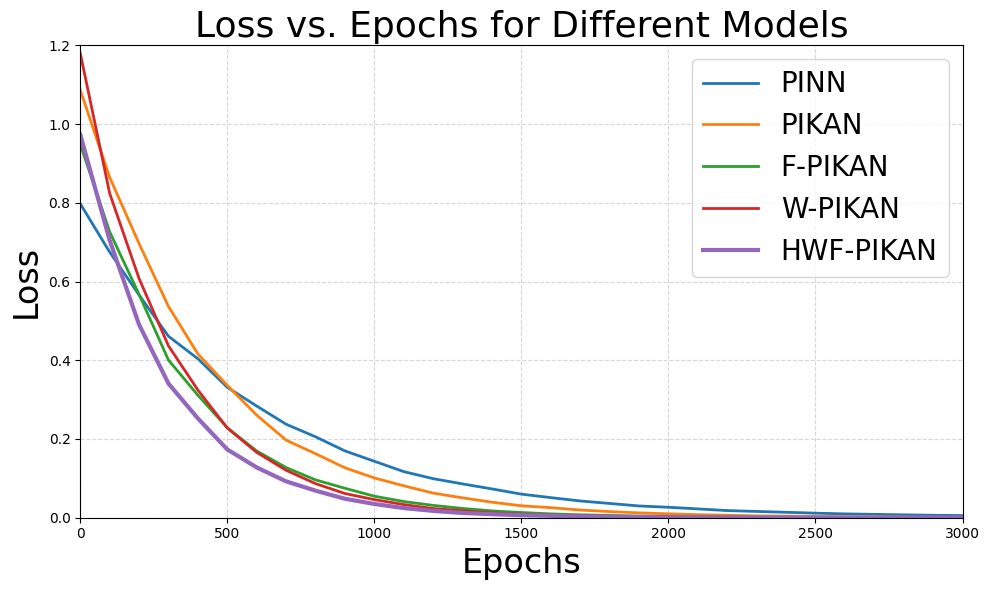}}
	\subfloat [Ablation Study (Case 2)]{\includegraphics[width=0.46\textwidth]{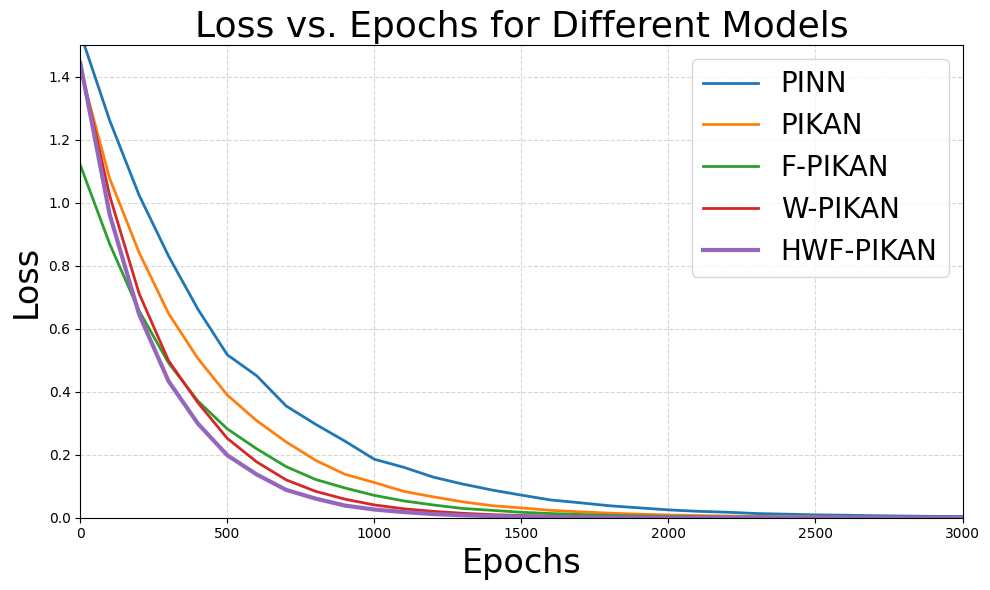}}
	\caption{Different analyses for 1D advection equation (continuous and discontinuous).} 
	\label{fig:6}
\end{figure}

\noindent  Furthermore, a sensitivity analysis on two key parameters of the HWF-PIKAN method, including the number of collocation points (NCP) and the number of epochs (NE) has been carried out. As shown in Figs.~\ref{fig:6} (a-b), more than 100 collocation points are needed to achieve a reasonable approximation for the continuous problem. Increasing NCP beyond this minimum can reduce computation time up to a certain threshold, after which both time and cost rise. Similarly, as depicted in Figs.~\ref{fig:6} (c-d), convergence to an accurate discontinuous solution required more than 10,000 epochs. These results emphasize the need to balance computational cost and accuracy when tuning NCP and NE. For the 1D problems shown in Fig.~\ref{fig:5} (a-g), 10,000 NCP and 20,000 NE were set to obtain much higher accuracy.

\noindent In general, the performance of any numerical method in terms of accuracy can be measured by evaluation of its error associated with the solver. In order to conduct an ablation study, the above problems have been solved via Fourier-enhanced PIKAN (F-PIKAN), wavelet-enhanced PIKAN (W-PIKAN), vanilla versions of PINN and PIKAN, in addition to HWF-PIKAN. Figs.~\ref{fig:6} (e-f) give comparison of the loss function (MSE error) with respect to the number of epochs for the above 1D problems.\\

\begin{figure}[H]
	\centering
	\subfloat [NCP=75]{\includegraphics[width=0.23\textwidth]{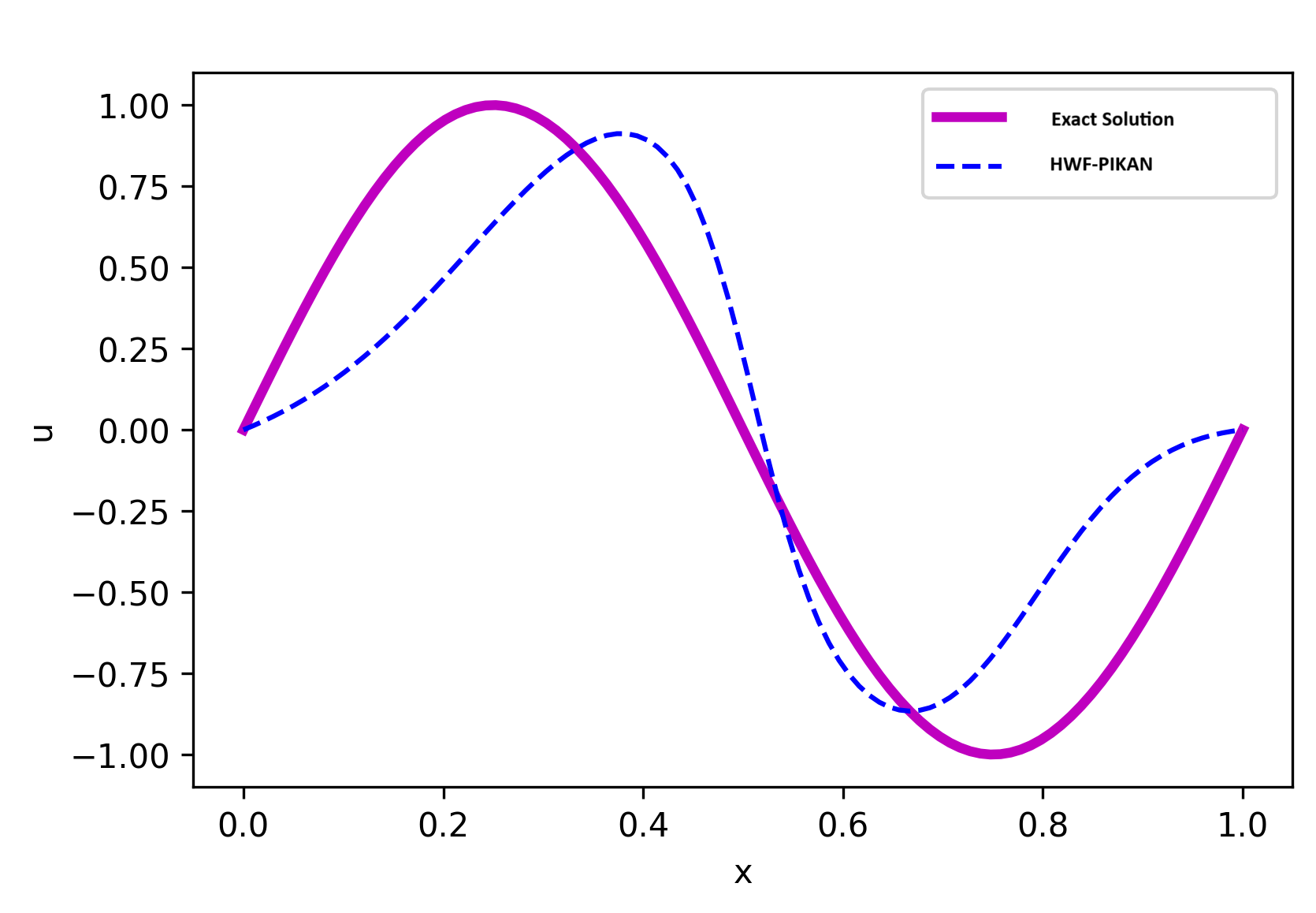}}
	\label{subfig:a} 
	\subfloat [NCP=100]{\includegraphics[width=0.23\textwidth]{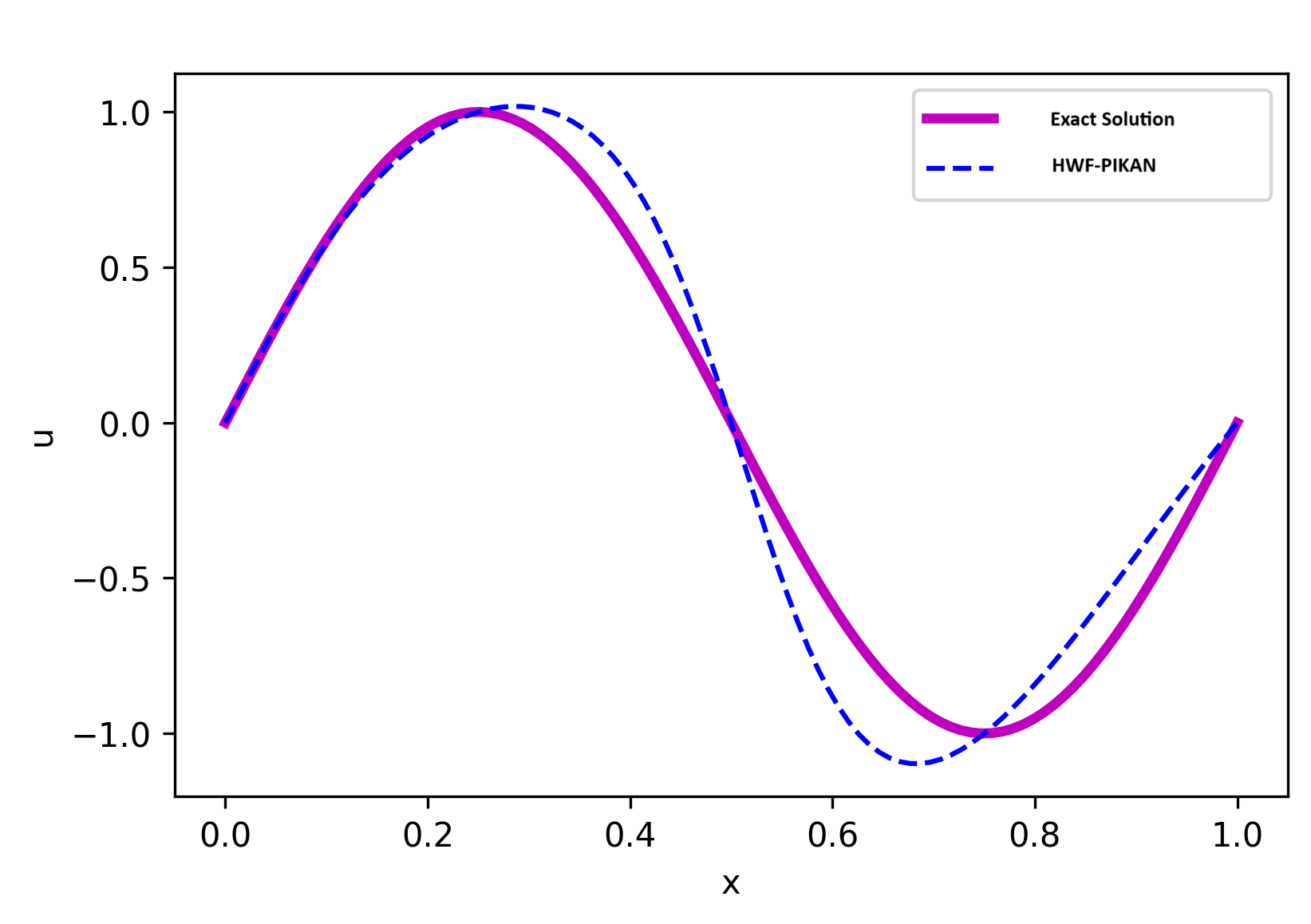}}
	\subfloat [NE=5000]{\includegraphics[width=0.23\textwidth]{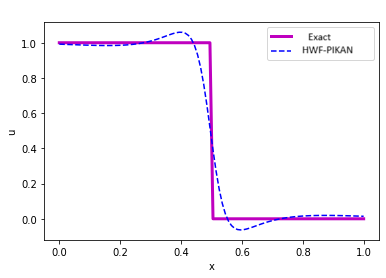}}
	\subfloat [NE=10000]{\includegraphics[width=0.23\textwidth]{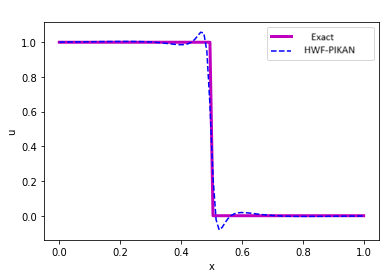}}
\end{figure}

\begin{figure}[H]	
	\subfloat [L2 error vs. training points]{\includegraphics[width=0.5\textwidth]{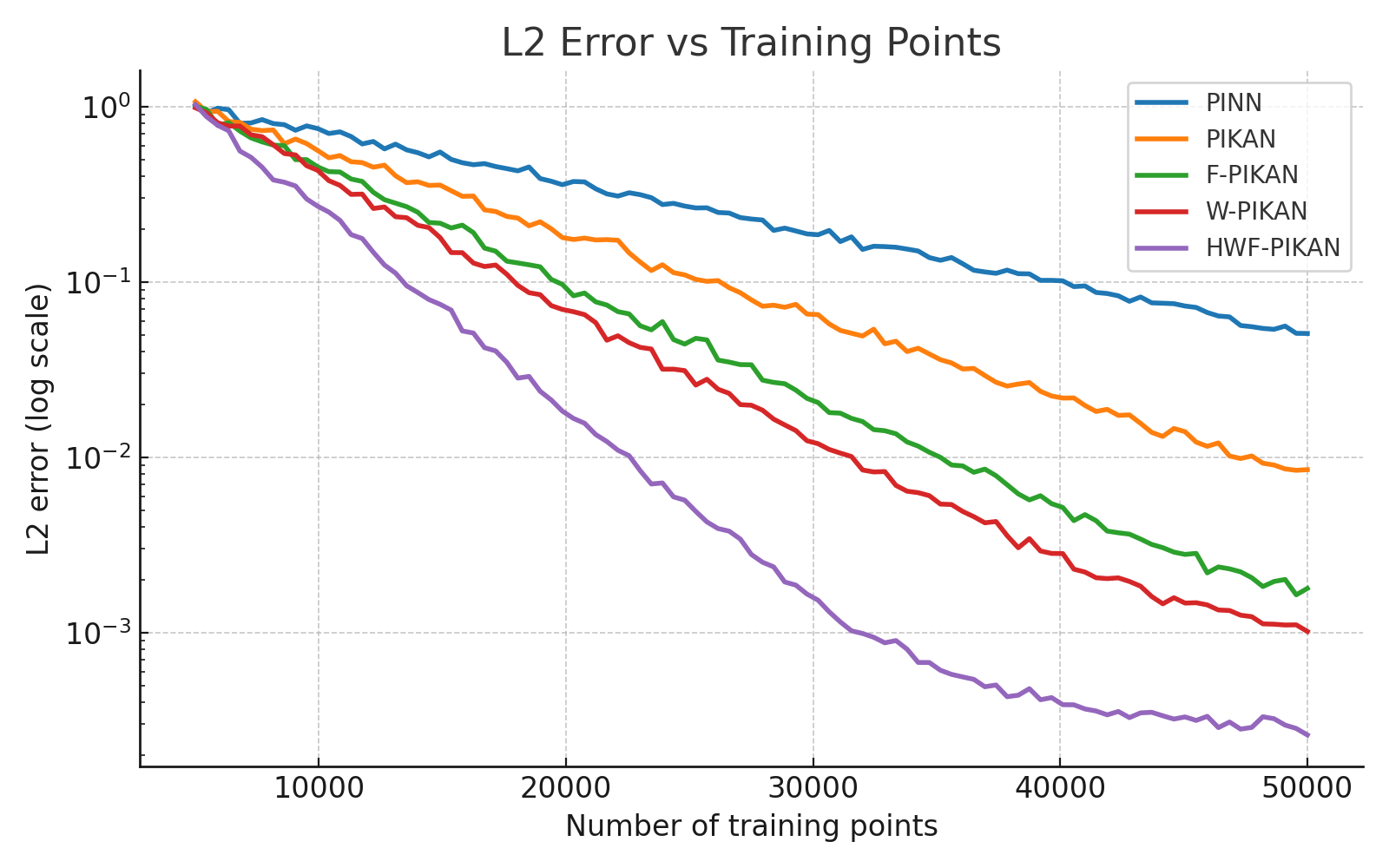}}
	\subfloat [L2 error vs. wall-clock time]{\includegraphics[width=0.5\textwidth]{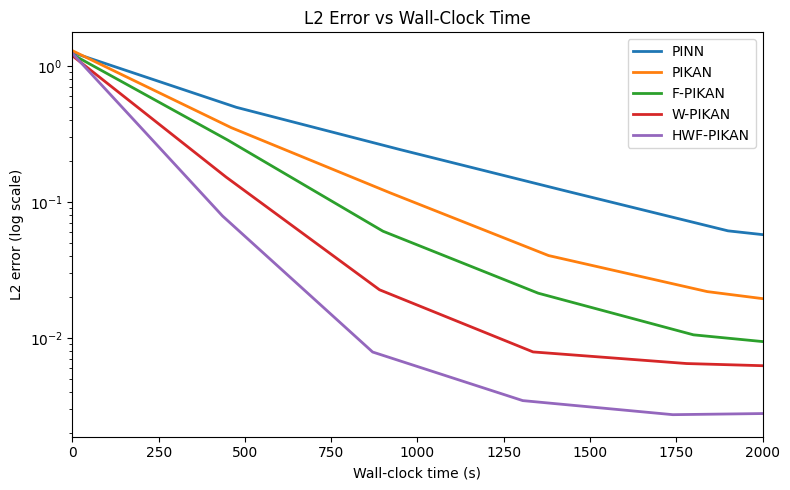}}
	\caption{Ablation study for 2D advection equation (Case 4).}
	\label{fig:9}
\end{figure}	

\noindent Another ablation study is conducted for the above 2D problems to show the superiority of the HWF-PIKAN solvers among others, which is illustrated in Fig.~\ref{fig:9}. The left subfigure highlights the convergence behavior of the method with respect to the number of collocation points, whereas the right subfigure reflects its computational efficiency by relating the error to the wall–clock time.

\noindent Table \ref{tab:1} summarizes all the configurations of the HWF-PIKAN model for the solved problems in space-time using PyTorch.

\begin{table}[H]
	\centering
	\caption{The space-time DF-PINN Configurations for the solved problems using PyTorch.}
	\renewcommand{\arraystretch}{1.5} 
	\resizebox{\textwidth}{!}{
		\begin{tabular}{|c|c|c|c|c|c|c|}
			\hline
			\textbf{Problem} & \textbf{NCP} & \textbf{LR} & \textbf{$\pmb{\phi} \, \textbf{(learnable)}$} & {$\mathbf{M \times N}$} & \textbf{Optimizer} & \textbf{Inputs} \\ \hline
			Case 1 (1D sinusoidal) & $10^{4}$ (Uniform) & $10^{-2}$ & Cubic B-Spline & $3 \times 15$ & Adam & \( x, t \) \\ \hline
			Case 2 (1D heaviside) & $10^{4}$ (Random) & $10^{-2}$ & Cubic B-Spline & $3 \times 15$ & Adam & \( x, t \) \\ \hline
			Case 3 (2D Gaussian) & $2\times 10^{5}$ (Random) & $10^{-3}$ & Cubic B-Spline & $4 \times 15$ & Adam + L-BFGS & \( x, y, t \) \\ \hline
			Case 4 (2D heaviside) & $3 \times 10 ^{5}$ (Random) & $10^{-3}$ & Cubic B-Spline & $4 \times 15$ & Adam + L-BFGS & \( x, y, t \) \\ \hline
		\end{tabular}
	}
	\label{tab:1}
\end{table}

\noindent The abbreviations used in table \ref{tab:1} are as follows: NCP: Number of Collocation Points; LR: Learning Rate; $\sigma$: Activation (Transfer) Function; $M \times N$: Layers $\times$ Neurons per Layer.

\subsection{Collisionless Boltzmann Equation (CBE) in Phase-Space}
Akin to spatial space in Lagrangian mechanics and Hilbert space in quantum mechanics, the concept of phase space density comes from Hamiltonian mechanics. This is based upon the idea that knowing the probability distribution of position and momentum of a system's particles allows for the calculation of all macroscopic properties. From this perspective, the dynamic state of a particle at a specific moment can be described by spatio-veloci-temporal vectors in a seven-dimensional phase-space ($x,y,z,v_{x},v_{y},v_{z},t$) a.k.a. $\mu$-space, in which each point corresponds to the dynamic state of that particle \cite{Boltzmann1}. Similarly, the dynamic state of a $N$-particle system can also be depicted by $N$ representative points in the $\mu$-space. The CBE (Vlasov equation), as a specific instance of Liouville's theorem, takes the form of a time-reversible advection equation in phase-space \cite{Vlasov2}. This equation with no force field term describes a system of $N$ free-streaming particles without collisions, or with very rare collisions, delineating the invariance of the phase-space density ($f$) along trajectories in $\mu$-space. This equation characterizes the dynamics of a continuous, incompressible phase fluid, represented by the mass density function $f(\mathbf{r}, \mathbf{p} , t)$. Unlike the collisional Boltzmann equation, the CBE does not account for random binary collisions. These collisions are what cause physical dissipation and irreversibility in a system \cite{Vlasov3}. \\

The full 7D (spatially-3D) and free-streaming 3D (spatially-1D) collisionless Boltzmann equations are as relations \eqref{eq:42} and \eqref{eq:43} respectively \cite{Vlasov4, Vlasov5}: 

\begin{equation}
	\frac{\partial{f(\mathbf{r}, \mathbf{p}, t)}}{\partial{t}} + \frac{\partial \mathbf{r}}{\partial t}\cdot \frac{\partial f(\mathbf{r}, \mathbf{p}, t)}{\partial \mathbf{r}} + \frac{d\mathbf{p}}{dt} \cdot \frac{\partial f(\mathbf{r}, \mathbf{p}, t)}{\partial \mathbf{p}} = 0
	\label{eq:42}	
\end{equation} 

\begin{equation}
	\frac{\partial{f(x, v_{x}, t)}}{\partial{t}} + v_{x} \frac{\partial{f(x, v_{x}, t)}}{\partial{x}} = 0 
	\label{eq:43}
\end{equation}	
	
\subsubsection{Macroscopic Properties Derived from the Distribution Function}

In essence, the aim of kinetic theory in a $N$-particle system is to predict the macroscopic behavior of the system from its microscopic information \cite{Boltzmann2}. For instance, a $N$-particle system could represent fundamental particles like electrons, phonons, photons, particles in a continuous or dilute gas, or even celestial bodies (e.g., interplanetary, interstellar, or dark matter) within a galaxy \cite{Vlasov1}. The macroscopic properties of a collisionless system or gas, governed by the collisionless Boltzmann equation, are obtained as velocity-space moments of the one-particle distribution function \( f(x,v,t) \). These properties describe the collective behavior of the system by averaging over the microscopic dynamics of particles. The macroscopic quantities are computed as moments of the microscopic distribution function. Given a distribution function \( f(x,v,t) \), which represents the probability density of finding a particle at time \( t \), position \( x \), and velocity \( v \), the macroscopic properties are computed as follows:

\begin{equation}
	n(x,t) = \int_{-\infty}^{\infty} f(x,v,t) \, dv,
\end{equation}

\begin{equation}
	u(x,t) = \frac{1}{n(x,t)} \int_{-\infty}^{\infty} v \, f(x,v,t) \, dv,
\end{equation}

\begin{equation}
	p(x,t) = \int_{-\infty}^{\infty} (v - u(x,t))^2 f(t,x,v) \, dv.
\end{equation}

\begin{equation}
	T(x,t) = \frac{1}{n(x,t)} \int_{-\infty}^{\infty} (v - u(x,t))^2 f(x,v,t) \, dv,
\end{equation}

\noindent In which, the number density \( n(x,t) \) indicates how many particles are present at a given point in space and time. The mean velocity \( u(x,t) \) represents the average motion of the particles, often interpreted as the macroscopic flow velocity. The pressure \( p(x,t) \) is associated with the momentum flux and random motion of particles around the mean velocity. The temperature \( T(x,t) \) reflects the thermal energy of the system, related to the variance of particle velocities.

\subsubsection{Cubic and Cylindrical Waterbag Problems}
In this study, the 3D spatio-veloci-temporal (spatially-1D) CBE with no force field term with different 3D initial conditions in phase-space is solved using the HWF-PIKAN framework. Conventional numerical methods for solving the CBE in the literature have predominantly employed the discretization-based methods \cite{waterbag3,emfra}. The so-called Vlasovian incompressible flow in phase-space can be visualized by considering a homogeneous, single-phase fluid, where the initial density $f(x, v, 0)$ is uniform $f=\eta$ within a specific region and zero outside of it. Throughout the evolution governed by the CBE, the density function stays piecewise uniform, and the volume of the so-called 'waterbag' remains constant, though it deforms into increasingly finer filaments over time \cite{waterbag2}. \\


In the first test case, the initial distribution is considered as a cubic waterbag confined within the phase-space bounds $-2<x<2$ and $-1<v<1$, with a computational domain $-4<x<4$ and $-2<v<2$. Within this domain, the waterbag is a uniform distribution with constant density, so that the initial distribution in phase space is represented as a box-shaped region with the limits specified as equation \eqref{eq:45}.

\begin{equation}
	Case \,5:\
f(x, v_{x}, 0) = \left\{ \begin{matrix}
1\ \ \ \ \ \ if\ (-2 \leq x \leq 2) \quad \& \quad (-1 \leq v_{x} \leq 1) \\ \\
0\ \ \ \ \ \ \ \ \ \ \ \ \ \ \ \ \ \ \ \ \ \ \ \ \ \ \ \ \ \ \ \ \ \ \ \ \ \ \ \ \  otherwise \\
\end{matrix} \right.\  \\ 
\label{eq:45}
\end{equation}

\noindent In the next case, the waterbag is initialized as a Cylindrical distribution in phase-space, defined by a radius of 2, with the computational domain as $-4<x<4$ and $-4<v<4$ as written in relation \eqref{eq:46}. 


\begin{equation}
	Case \,6:\
	f(x, v_{x}, 0) = \left\{ \begin{matrix}
		1\ \ \ \ \ \ \ \ \ \ \ if\ (x^2 + v_{x}^2) \leq 4 \\ \\
		0\ \ \ \ \ \ \ \ \ \ \ \ \ \ \ \ \ \ \  otherwise \\
	\end{matrix} \right.\  \\ 
	\label{eq:46}
\end{equation}

\begin{figure}[H]
\centering
\subfloat [Case 5 (t=0)]{\includegraphics[width=0.22\textwidth]{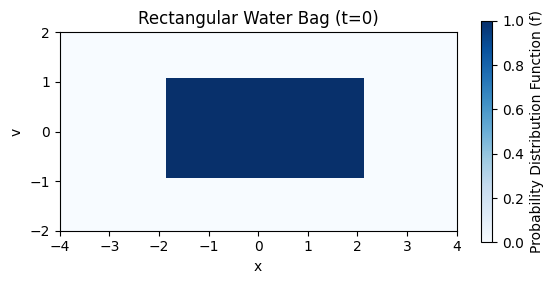}} 
\subfloat [Case 5 (t=1]{\includegraphics[width=0.22\textwidth]{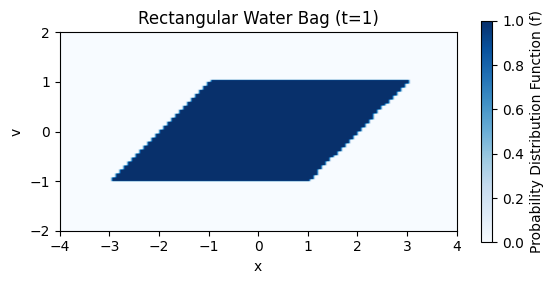}} 
\subfloat [Case 5 (t=3]{\includegraphics[width=0.22\textwidth]{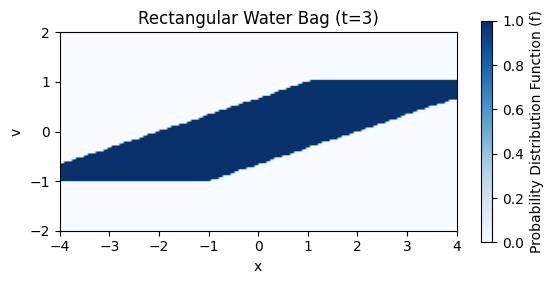}}
\subfloat [Case 5 (t=4]{\includegraphics[width=0.22\textwidth]{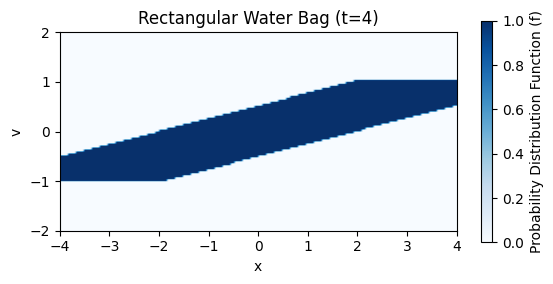}}
\\
\subfloat [Case 5 (t=0]{\includegraphics[width=0.22\textwidth]{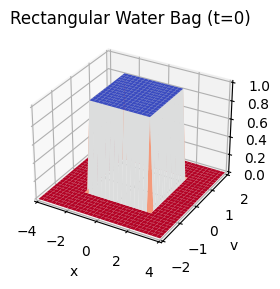}}
\subfloat [Case 5 (t=1)]{\includegraphics[width=0.22\textwidth]{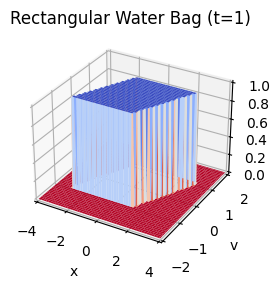}}
\subfloat [Case 5 (t=3)]{\includegraphics[width=0.22\textwidth]{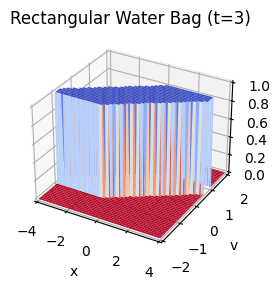}}
\subfloat [Case 5 (t=4)]{\includegraphics[width=0.22\textwidth]{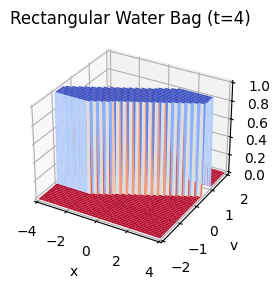}}
\\
\subfloat [Case 6 (t=0)]{\includegraphics[width=0.22\textwidth]{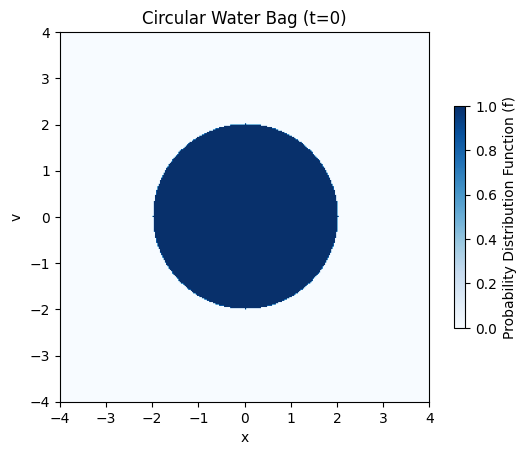}} 
\subfloat [Case 6 (t=1)]{\includegraphics[width=0.22\textwidth]{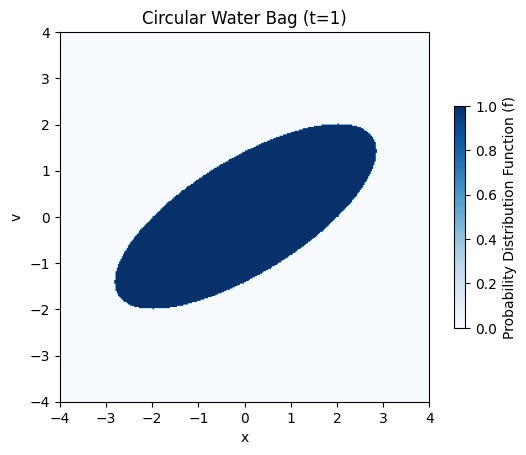}} 
\subfloat [Case 6 (t=2)]{\includegraphics[width=0.22\textwidth]{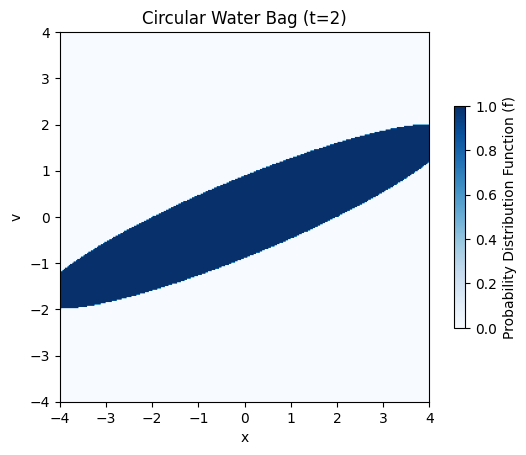}}
\subfloat [Case 6 (t=4)]{\includegraphics[width=0.22\textwidth]{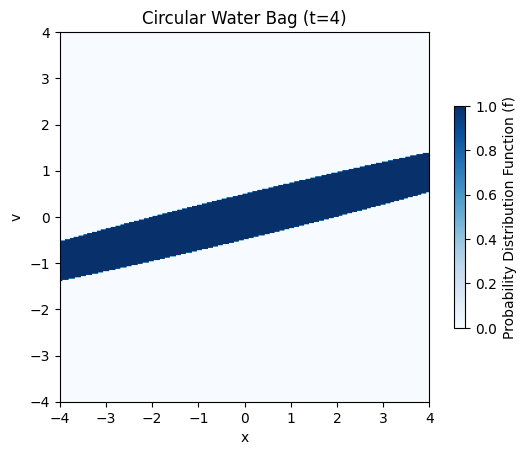}}
\\
\subfloat [Case 6 (t=0)]{\includegraphics[width=0.22\textwidth]{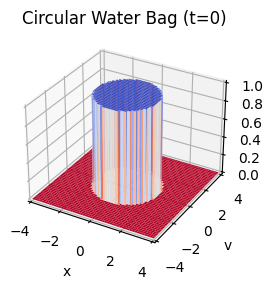}}
\subfloat [Case 6 (t=1)]{\includegraphics[width=0.22\textwidth]{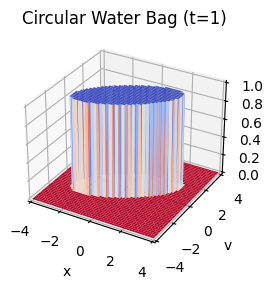}}
\subfloat [Case 6 (t=2)]{\includegraphics[width=0.22\textwidth]{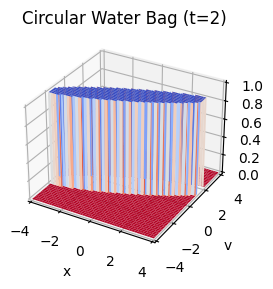}}
\subfloat [Case 6 (t=4)]{\includegraphics[width=0.22\textwidth]{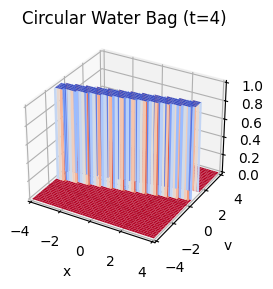}}
\caption{HWF-PIKAN solutions for CBE with a discontinuous, 3D cubic and cylindrical waterbag ICs.}
\label{fig:11}
\end{figure}

\begin{figure}[H]
	\centering
	\subfloat [Density (Case 5)]{\includegraphics[width=0.45\textwidth]{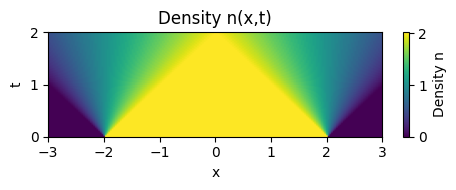}}
	\subfloat [Mean Velocity (Case 5)]{\includegraphics[width=0.45\textwidth]{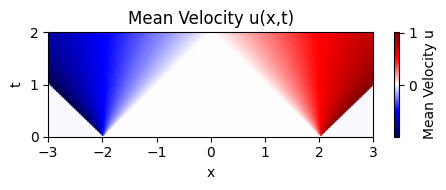}}
	\\
	\subfloat [Pressure (Case 5)]{\includegraphics[width=0.45\textwidth]{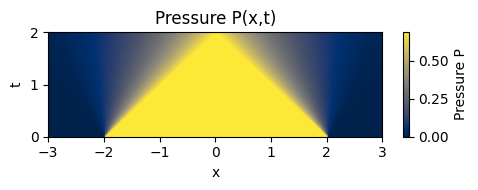}}
	\subfloat [Temperature (Case 5)]{\includegraphics[width=0.45\textwidth]{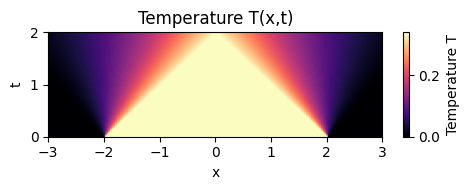}}
	\\
	\subfloat [Density (Case 6)]{\includegraphics[width=0.45\textwidth]{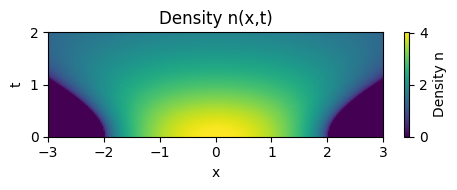}}
	\subfloat [Mean Velocity (Case 6)]{\includegraphics[width=0.45\textwidth]{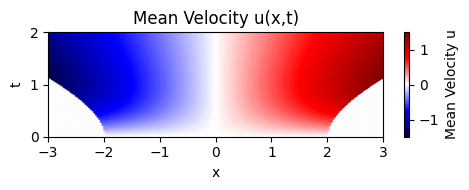}}
	\\
	\subfloat [Pressure (Case 6)]{\includegraphics[width=0.45\textwidth]{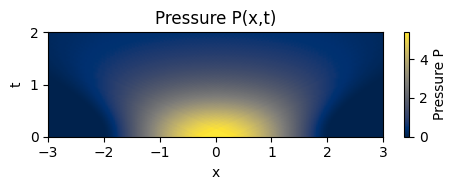}}
	\subfloat [Temperature (Case 6)]{\includegraphics[width=0.45\textwidth]{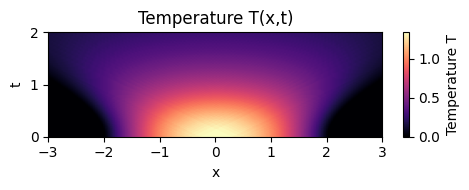}}
	\caption{Calculated macroscopic properties for cases 5 and 6.}
	\label{fig:12}		
\end{figure}

\noindent As time progresses, the CBE describes how the particles inside the waterbag evolves due to free streaming, as shown in Fig.~\ref{fig:11}. The particles will spread out in space over time, mimicking the behavior of a gas expanding into a vacuum \cite{freestream}. In the absence of collisions, the velocity distribution remains unchanged, but the particles will start spreading out spatially as they move. The waterbag will deform into increasingly finer structures, creating elongated filaments. This phenomenon is known as "filamentation". These filaments form as a result of differences in the velocities and spatial displacements of the particles over time \cite{waterbag3}. For instance, a highly rarefied gas in a free-molecular regime (with Knudsen number larger than 10) can be simulated by such collisionless system governed by the CBE \cite{freestream}. Macroscopic properties for such a gas in free-moleculare regime are calculated as presented in Fig.~\ref{fig:12}.\\

\subsubsection{Gaussian and Maxwellian Distributions}

For the next case, instead of a piece-wise waterbag, the initial condition is considered as a Gaussian distribution in phase-space, as stated in relations \eqref{eq:47} and \eqref{eq:48}:   
	
\begin{equation}
	Case \,7:\
f(x,v_{x},0) = \alpha \, \exp [\beta(x^{2} + v_{x}^{2})]
	\label{eq:47}
\end{equation}	

\begin{equation}
assuming:	\alpha=\frac{1}{2 \pi \sigma_{x} \sigma_{v_{x}}}, \quad \quad \beta=\frac{-1}{2(\sigma_{x}^2 + \sigma_{v_{x}}^2)}
	\label{eq:48}
\end{equation}
\\

\noindent This represents a smooth distribution where the number density $(f)$ decays radially from the origin at $(x, v_{x})=(0, 0)$. As the Gaussian distribution is spanned over both position and velocity, this distribution will also deform through time evolution. Since the number density is not piece-wise here, the new density of each spatio-temporal point should be calculated based on new distributions at a specific moment. \\

\begin{figure}[H]
\centering
\subfloat [Case 7 (t=0)]{\includegraphics[width=0.22\textwidth]{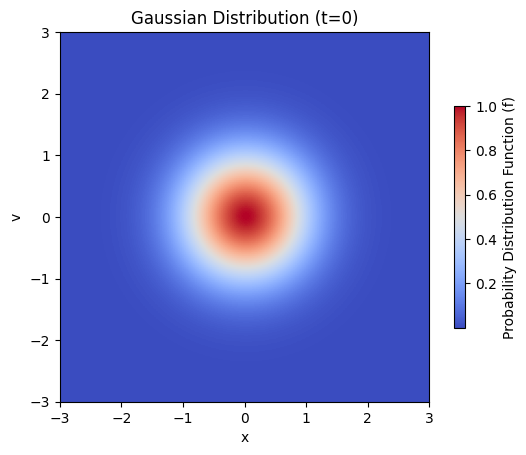}} 
\subfloat [Case 7 (t=1)]{\includegraphics[width=0.22\textwidth]{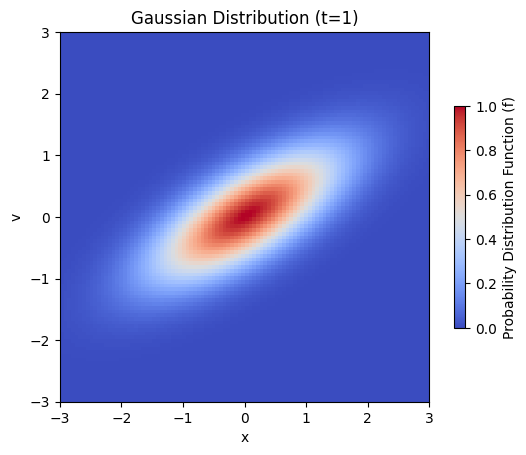}} 
\subfloat [Case 7 (t=2)]{\includegraphics[width=0.22\textwidth]{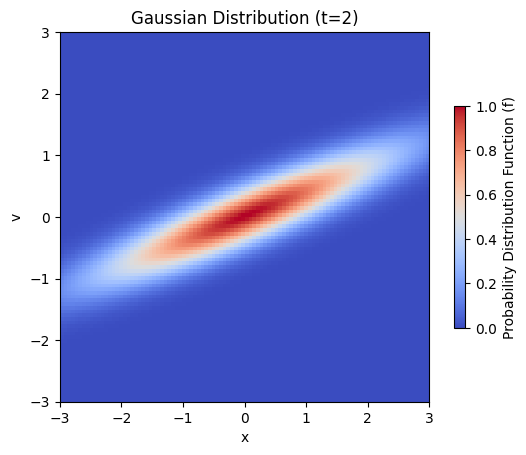}}
\subfloat [Case 7 (t=4)]{\includegraphics[width=0.22\textwidth]{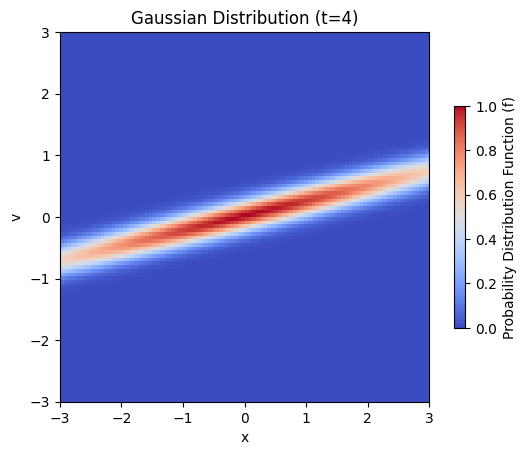}}
\\
\subfloat [Case 7 (t=0)]{\includegraphics[width=0.22\textwidth]{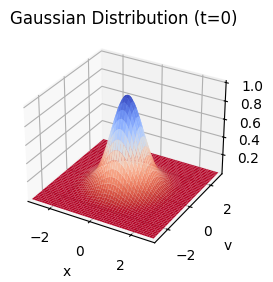}}
\subfloat [Case 7 (t=1)]{\includegraphics[width=0.22\textwidth]{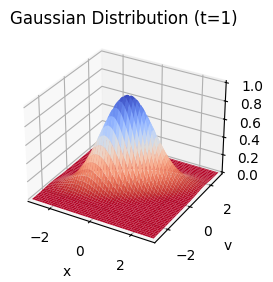}}
\subfloat [Case 7 (t=2)]{\includegraphics[width=0.22\textwidth]{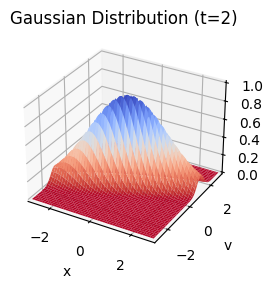}}
\subfloat [Case 7 (t=4)]{\includegraphics[width=0.22\textwidth]{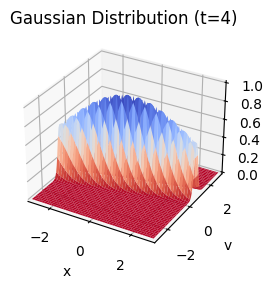}}
\\
\subfloat [Case 8 (t=0)]{\includegraphics[width=0.22\textwidth]{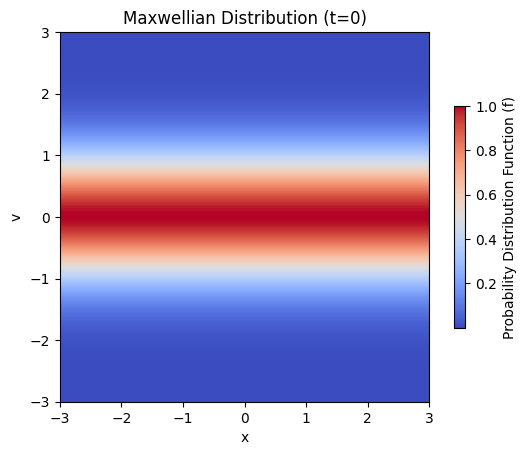}} 
\subfloat [Case 8 (t=2)]{\includegraphics[width=0.22\textwidth]{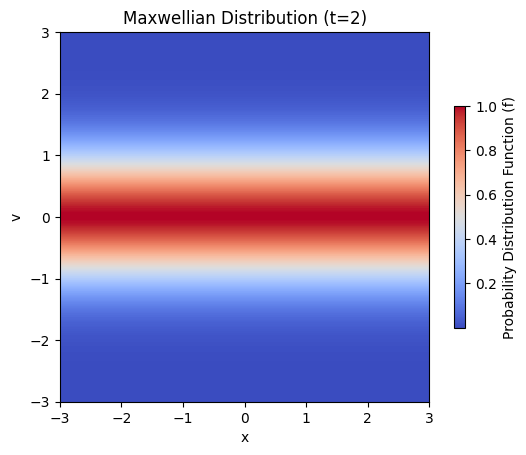}} 
\subfloat [Case 9 (t=0)]{\includegraphics[width=0.22\textwidth]{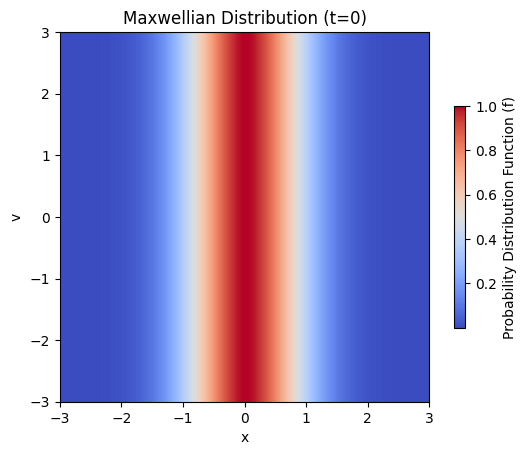}}
\subfloat [Case 9 (t=2)]{\includegraphics[width=0.22\textwidth]{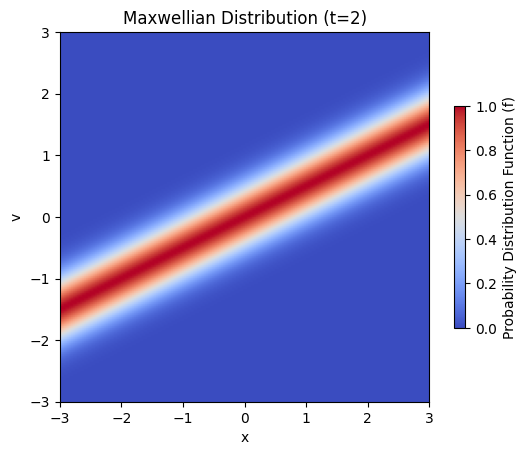}}
\\
\subfloat [Case 8 (t=0)]{\includegraphics[width=0.22\textwidth]{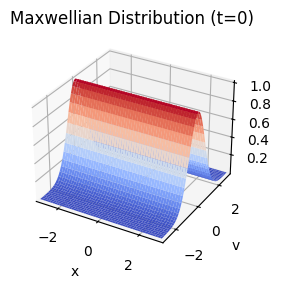}}
\subfloat [Case 8 (t=2)]{\includegraphics[width=0.22\textwidth]{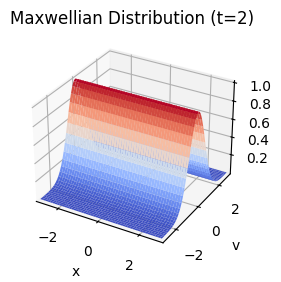}}
\subfloat [Case 9 (t=0)]{\includegraphics[width=0.22\textwidth]{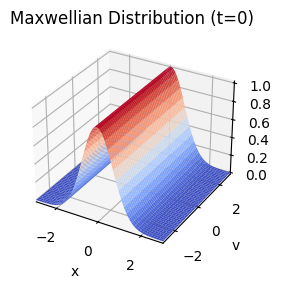}}
\subfloat [Case 9 (t=2)]{\includegraphics[width=0.22\textwidth]{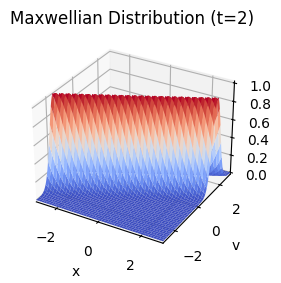}}
\caption{HWF-PIKAN solutions for CBE with a continuous, 3D Gaussian and Maxwellian distributions ICs.}
\label{fig:13}
\end{figure}

On the other hand, the Maxwellian distribution might be used as an initial condition for the distribution function in a collisionless system, especially when the system is expected to have some properties similar to those in thermal equilibrium. For example, in some collsionless systems, the distribution function might evolve towards a quasi-equilibrium states. In the next cases, the initial conditions are Maxwell-Boltzmann (Maxwellian) distributions over velocities and spatial space, as written in relations \eqref{eq:49} and \eqref{eq:492} respectively:

\begin{equation}
	Case \,8:\
	f(x,v_{x},0) = \alpha \, \exp (\beta v_{x}^{2})
	\label{eq:49}
\end{equation}	

\begin{equation}
	Case \,9:\
	f(x,v_{x},0) = \alpha \, \exp (\beta x^{2})
	\label{eq:492}
\end{equation}	

\begin{equation}
	assuming: \quad	\alpha=\frac{1}{\sqrt{2 \pi \sigma^2}} \, , \quad \quad \beta=\frac{-1}{2\sigma^2}
	\label{eq:50}
\end{equation}

\noindent where $\sigma$ is the standard deviation of the velocity and spatial distributions. As depicted in Figs.~\ref{fig:13} (a-h), the initially circular Gaussian distribution will elongate, transitioning into an elliptical shape in phase space. As the evolution progresses, the initial smoothness will be lost, and the phase-space distribution will develop sharp, thin filaments along certain directions in phase-space, which in turn indicates the onset of filamentation due to the differential velocities of the particles in the system.\\

Generally, the CBE does not single out the Maxwell-Boltzmann (Maxwellian) distribution. The effect of binary collisions must be included to ensure that the distribution will relax to a Maxwellian distribution. However, it can still be considered as an initial condition for the CBE over velocities and spatial space. As shown in Figs.~\ref{fig:13} (i–j, m–n), the absence of spatial dependence in the velocity distribution allows it to remain Maxwellian \cite{relax1}. In this case, the macroscopic properties remain unchanged. By contrast, when the Maxwellian distribution varies across space, the phase-space density evolves over time, leading to a change in its shape, as depicted in Figs.~\ref{fig:13} (k-l, o-p). \\	

\begin{figure}[H]
	\centering
	\subfloat [Density (Case 7)]{\includegraphics[width=0.45\textwidth]{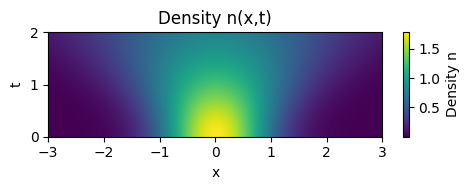}}
	\subfloat [Mean Velocity (Case 7)]{\includegraphics[width=0.45\textwidth]{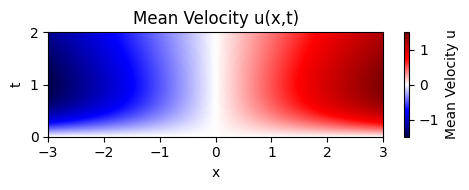}}
	\\
	\subfloat [Pressure (Case 7)]{\includegraphics[width=0.45\textwidth]{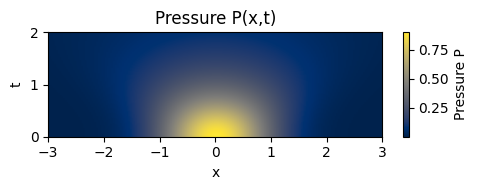}}
	\subfloat [Temperature (Case 7)]{\includegraphics[width=0.45\textwidth]{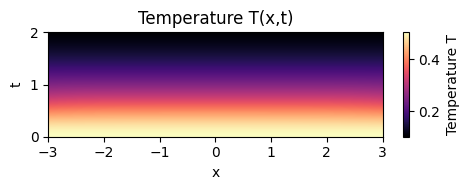}}
	\\
	\subfloat [Density (Case 9)]{\includegraphics[width=0.45\textwidth]{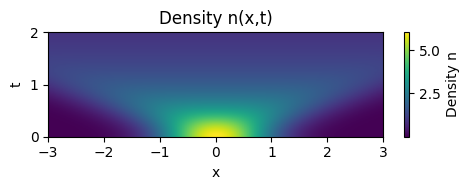}}
	\subfloat [Mean Velocity (Case 9)]{\includegraphics[width=0.45\textwidth]{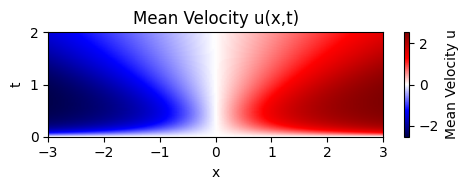}}
	\\
	\subfloat [Pressure (Case 9)]{\includegraphics[width=0.45\textwidth]{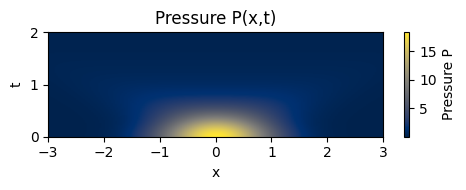}}
	\subfloat [Temperature (Case 9)]{\includegraphics[width=0.45\textwidth]{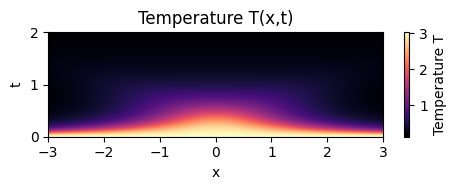}}	
	\caption{Calculated macroscopic properties for cases 7 and 9.}
	\label{fig:14}		
\end{figure}

\noindent It is worth noting that even in a collisionless system, the concept of temperature can still be used to characterize the average kinetic energy of the particles, and the distribution function may exhibit some features similar to the Maxwellian distribution \cite{relax1}. For example, the distribution might peak around a certain velocity, and the width of the distribution could be related to the temperature of the system. Therefore, all the macroscopic quantities can also be computed as shown in Fig.~\ref{fig:14}. It should be noted that, in case 12, the macroscopic properties remain constant in both time and space, since the variation of the distribution function with respect to x is zero.  \\

\subsubsection{Sod Shock Tube in Highly Rarefied Regime via CBE}

The Sod shock tube problem is a classical test case for compressible flow, frequently used to validate numerical methods for hyperbolic conservation laws. In this study, we solve the problem using the collisionless Boltzmann equation (CBE). The absence of a collision term corresponds to the free molecular (ballistic motion) regime, in which the Knudsen number $\mathrm{Kn} > 10$, so that intermolecular collisions are negligible and particle transport is dominated by free-streaming. The initial condition is a discontinuous Maxwellian distribution corresponding to two gas states separated at $x = 0.5$ as relation \eqref{eq:51}.


\begin{equation}
	Case \,10:\ 
	f(x,v,0) =
	\begin{cases} 
		\dfrac{\rho_L}{\sqrt{2\pi T_L}} \exp\Big[-\dfrac{(v-u_L)^2}{2 T_L}\Big], & x < 0.5, \\ \\
		\dfrac{\rho_R}{\sqrt{2\pi T_R}} \exp\Big[-\dfrac{(v-u_R)^2}{2 T_R}\Big], & x \ge 0.5,
	\end{cases}
	\label{eq:51}
\end{equation}
\\
\noindent where $(\rho_L, u_L, T_L) = (1.0, 0.0, 1.0)$ and $(\rho_R, u_R, T_R) = (0.125, 0.0, 0.8)$ are the density, mean velocity, and temperature of the left and right regions, respectively.\\

\begin{figure}[H]
	\centering
	\subfloat [Density (Case 10)]{\includegraphics[width=0.45\textwidth]{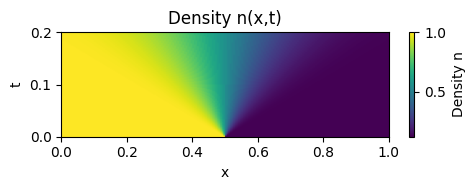}}
	\subfloat [Mean Velocity (Case 10)]{\includegraphics[width=0.45\textwidth]{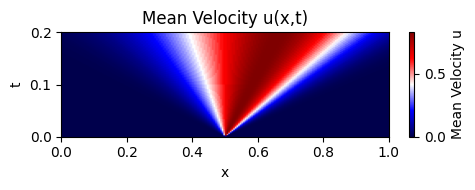}}
	\\
	\subfloat [Pressure (Case 10)]{\includegraphics[width=0.45\textwidth]{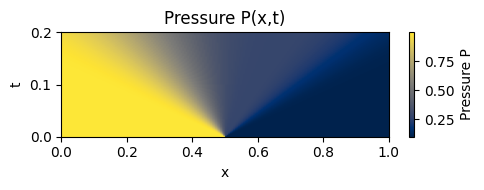}}
	\subfloat [Temperature (Case 10)]{\includegraphics[width=0.45\textwidth]{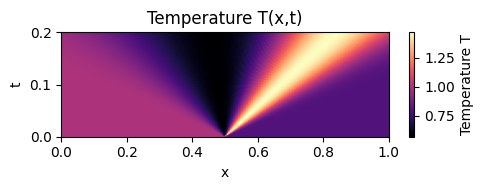}}
	\caption{Microscopic and macroscopic quantities for Sod problem (case 10).}
	\label{fig:15}
\end{figure}

In this case, the temporal evolution of the distribution function is illustrated in Figs.~\ref{fig:15} (a-d), and the corresponding macroscopic quantities are depicted in Figs.~\ref{fig:15} (e-l). The results demonstrate that the initially discontinuous shock waves become progressively smoother with time. This phenomenon is attributed to the relatively large Knudsen number, which is a defining characteristic of highly rarefied gas flows.

\begin{figure}[H]
	\centering
	\subfloat [Case 10 (t=0)]{\includegraphics[width=0.32\textwidth]{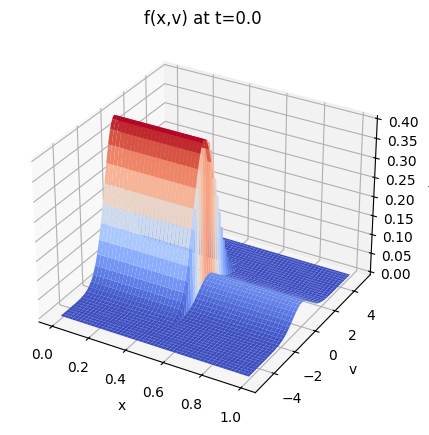}} 
	\hspace{1.5cm}
	\subfloat [Case 10 (t=0.2)]{\includegraphics[width=0.32\textwidth]{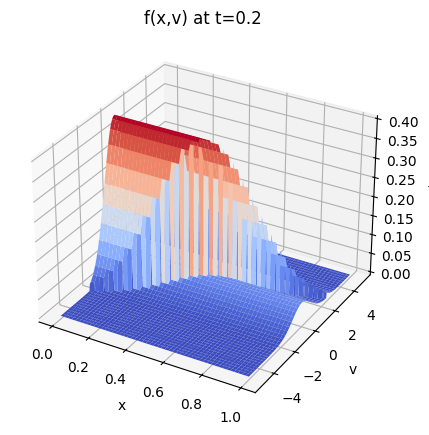}} 
	\\
	\subfloat [Case 10 (t=0)]{\includegraphics[width=0.45\textwidth]{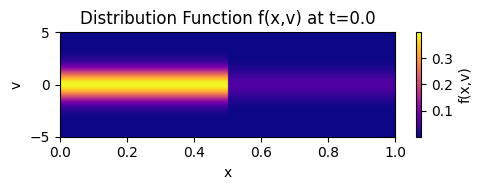}}
	\subfloat [Case 10 (t=0.2)]{\includegraphics[width=0.45\textwidth]{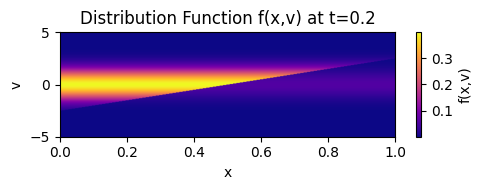}}
	\\
	\subfloat [Density (t=0.2)]{\includegraphics[width=0.225\textwidth]{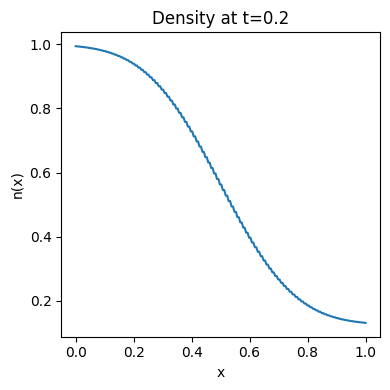}}
	\subfloat [Velocity (t=0.2)]{\includegraphics[width=0.225\textwidth]{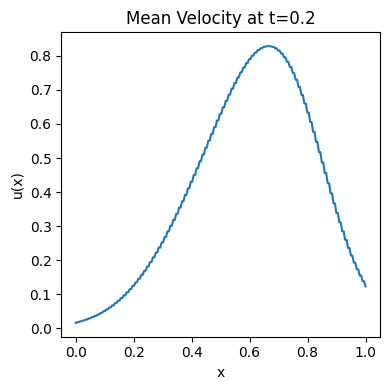}}
	\subfloat [Pressure (t=0.2)]{\includegraphics[width=0.225\textwidth]{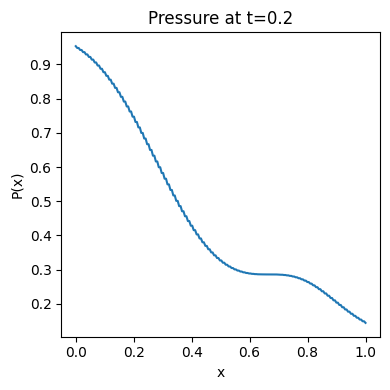}}
	\subfloat [Temperature (t=0.2)]{\includegraphics[width=0.225\textwidth]{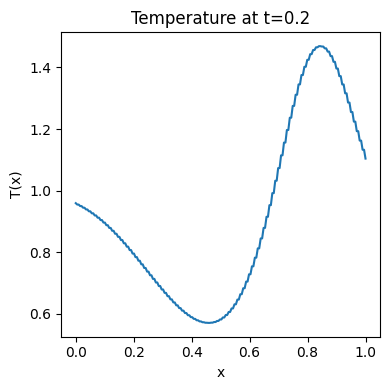}}
\end{figure}

\noindent Table~\ref{tab:3} summarizes all the configurations of phase-space HWF-PIKAN model for the solved problems based on collisionless Boltzmann equation.

\begin{table}[H]
	\centering
	\caption{The phase-space HWF-PIKAN Configurations for the solved problems.}
	\renewcommand{\arraystretch}{1.5} 
	\resizebox{\textwidth}{!}{
		\begin{tabular}{|c|c|c|c|c|c|c|}
			\hline
			\textbf{Problem} & \textbf{NCP} & \textbf{LR} & \textbf{$\pmb{\phi} \, \textbf{(learnable)}$} & {$\mathbf{M \times N}$} & \textbf{Optimizer} & \textbf{Inputs} \\ \hline
			Case 5 (Cubic Waterbag) & $3 \times 10^{5}$ (LHS) & $10^{-3}$ & Cubic B-Spline & $5 \times 15$ & Adam + L-BFGS & \( x, v_x, t \) \\ \hline
			Case 6 (Cylindrical Waterbag) & $3 \times 10^{5}$ (LHS) & $10^{-3}$ & Cubic B-Spline & $5 \times 15$ & Adam + L-BFGS & \( x, v_x, t \) \\ \hline
			Case 7 (Gaussian Distribution) & $3 \times 10^{5}$ (LHS) & $10^{-3}$ & Cubic B-Spline & $5 \times 15$ & Adam + L-BFGS & \( x, v_x, t \) \\ \hline
			Case 8 (Maxwellian over Velocity) & $3 \times 10^{5}$ (LHS) & $10^{-3}$ & Cubic B-Spline & $5 \times 15$ & Adam + L-BFGS  & \( x, v_x, t \) \\ \hline
			Case 9 (Maxwellian over Position) & $3 \times 10^{5}$ (LHS) & $10^{-3}$ & Cubic B-Spline & $5 \times 15$ & Adam + L-BFGS  & \( x, v_x, t \) \\ \hline
			Case 10 (Sod Shock Tube Problem) & $3 \times 10^{5}$ (LHS) & $10^{-3}$ & Cubic B-Spline & $5 \times 15$ & Adam + L-BFGS  & \( x, v_x, t \) \\ \hline
		\end{tabular}
	}
	\label{tab:3}
\end{table}

\noindent The abbreviation used in table~\ref{tab:3} is as follows: LHS: Latin Hypercube Sampling.\\	

\noindent Table~\ref{tab:2} reports the total training loss versus epochs for five models solving Sod problem via the CBE in phase space. The proposed HWF-PIKAN, combining hybrid wavelet–Fourier embeddings, reaches the lowest loss, demonstrating superior efficiency and accuracy in capturing phase-space dynamics.

\begin{table}[H]
	\centering
	\caption{Loss values vs. epochs for five models solving CBE in phase-space}
	\label{tab:loss_epochs}
	\begin{tabular}{|c|c|c|c|c|c|}
		\hline
		Epochs & Vanilla PINN & Vanilla PIKAN & F-PIKAN & W-PIKAN & HWF-PIKAN \\
		\hline
		100     & 1.23e+0 & 9.33e-1 & 1.68e+0 & 8.05e-1 & 7.07e-1 \\
		1000  & 4.20e-2 & 3.00e-2 & 2.80e-2 & 2.60e-2 & 2.40e-2 \\
		2000  & 2.30e-2 & 1.60e-2 & 1.40e-2 & 1.20e-2 & 1.00e-2 \\
		5000  & 8.50e-3 & 5.20e-3 & 4.50e-3 & 3.80e-3 & 3.20e-3 \\
		10000 & 3.20e-3 & 1.80e-3 & 1.40e-3 & 1.20e-3 & 9.00e-4 \\
		\hline
	\end{tabular}
	\label{tab:2}
\end{table}
	
\subsubsection{Lynden Bell's Theory of Violent Relaxation}
The concept of violent relaxation was introduced by Donald Lynden-Bell in 1967 to describe the initial rapid process by which a collisionless system of particles, such as galaxies, star clusters, charged particles in plasma or highly rarefied gaseous particles, reaches a quasi-stationary state (QSS) after a significant disturbance. According to his theory, based on the CBE, collisionless system undergoes a fast evolution initially, followed by a slower process. Essentially, violent relaxation describes this quick initial change until reaching a quasi-equilibrium state and the subsequent slower evolution afterwards \cite{relax6}.\\

\begin{figure}[H]
	\centering
	\subfloat [t=0]{\includegraphics[width=0.22\textwidth]{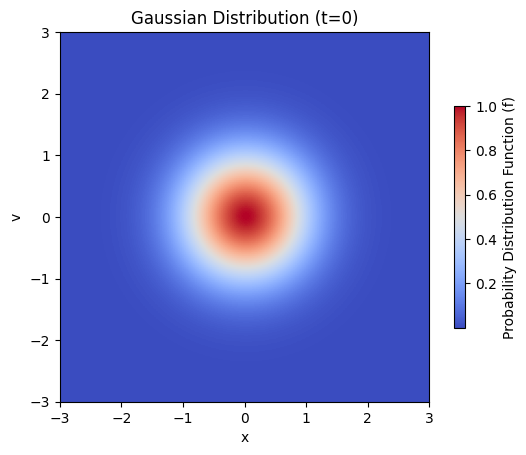}} 
	\subfloat [t=0.25]{\includegraphics[width=0.22\textwidth]{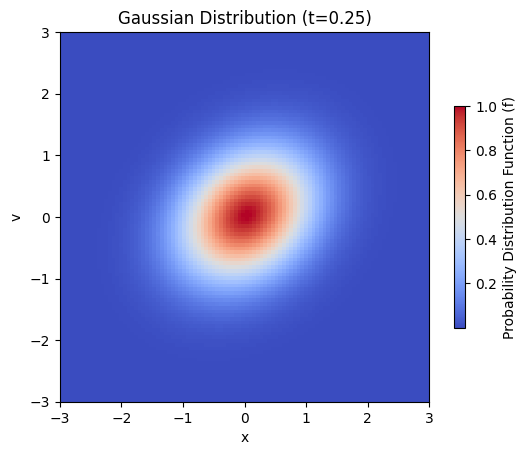}} 
	\subfloat [t=0.5]{\includegraphics[width=0.22\textwidth]{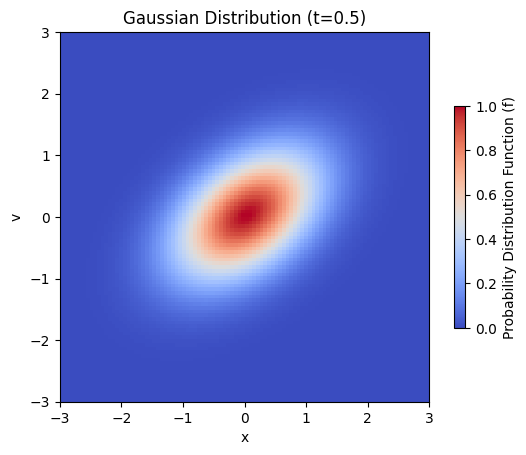}}
	\subfloat [t=0.75]{\includegraphics[width=0.22\textwidth]{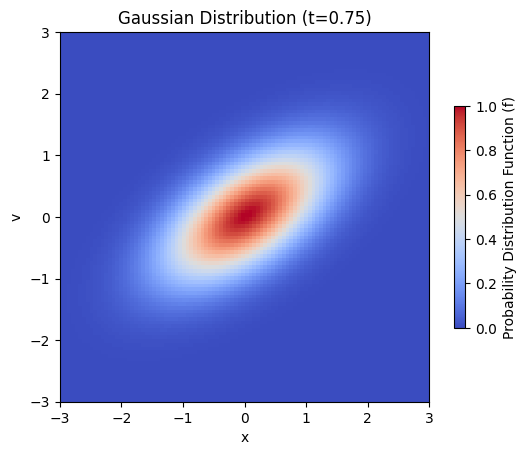}}
	\\
	\subfloat [t=3]{\includegraphics[width=0.22\textwidth]{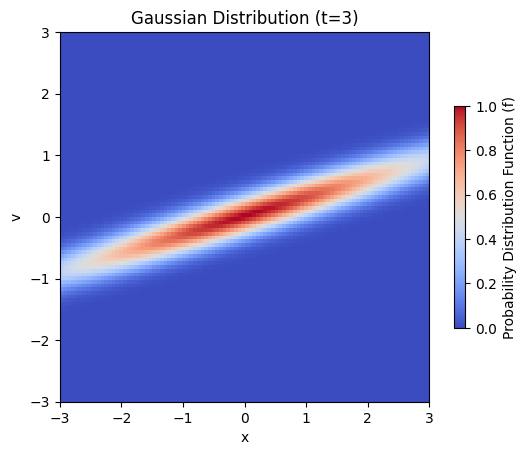}}
	\subfloat [t=3.25]{\includegraphics[width=0.22\textwidth]{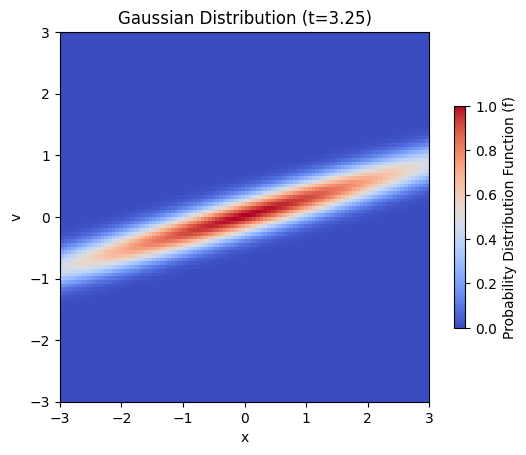}}
	\subfloat [t=3.5]{\includegraphics[width=0.22\textwidth]{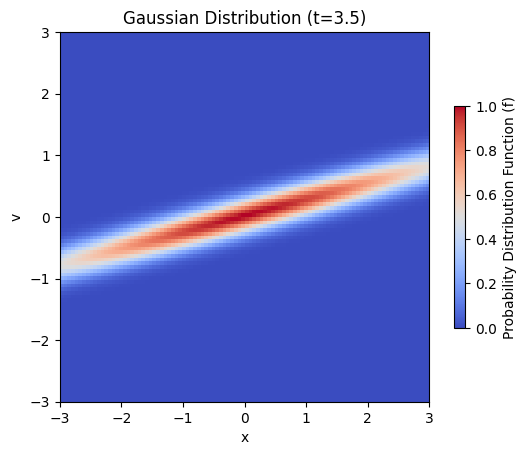}}
	\subfloat [t=3.75]{\includegraphics[width=0.22\textwidth]{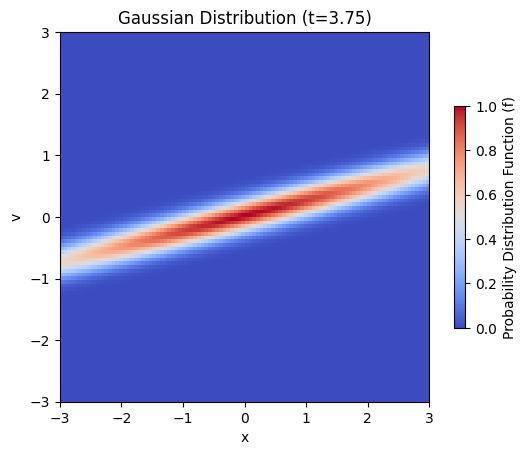}}
	\caption{Lynden Bell's theory of violent relaxation: figures a,b,c,d refer to so-called violent rapid relaxation, and figures e,f,g,h represent slow relaxation after reaching QSS.}
	\label{fig:16}
\end{figure}

As illustrated in Figs.~\ref{fig:16} (a-d), a significant "so-called violent" change in the phase-space initial condition is observed within the first second of temporal evolution ($t = 0,\, 0.25,\, 0.5,\, 0.75$). In contrast, a much slower change is noted by the fourth second of the evolution ($t = 3,\, 3.25,\, 3.5,\, 3.75$), as depicted in Figs.~\ref{fig:16} (e-h), indicating that significant changes will occur over extended periods following the quasi-stationary state (QSS). This phenomenon has been extensively discussed in the context of galactic dynamics and star formation. The theory is crucial for understanding the formation and evolution of galaxies and star clusters.

\section{Conclusion}
In this work, we developed a Hybrid Wavelet-Fourier-Enhanced Physics-Informed Kolmogorov-Arnold Network (HWF-PIKAN), a multi-resolution SciML framework for solving PDEs in a fully data-free setting. By combining local wavelet embeddings with global Fourier features, the proposed architecture enhances the representational capacity of physics-informed models, enabling accurate resolution of both smooth and discontinuous features while maintaining computational efficiency. Through systematic benchmarks on advection problems in one and two dimensions, including Cartesian, polar, and curvilinear coordinate systems, we demonstrated that HWF-PIKAN consistently outperforms baseline (vanilla) PINN, PIKAN, and their Fourier- and Wavelet-enhanced counterparts in both accuracy and convergence speed. Extending the framework to the collisionless Boltzmann (Vlasov) equation with a continuous-velocity formulation further confirmed the method’s adaptability to high-dimensional PDEs, offering a robust alternative to conventional discretization-based solvers. These results highlight the potential of multi-resolution spectral embeddings to advance physics-informed deep learning for complex kinetic and dynamical systems. Beyond advection and phase-space transport, the proposed approach provides a foundation for addressing a wide class of PDEs arising in fluid dynamics, plasma physics, astrophysics, and statistical mechanics.\\
\\

\noindent \textbf{Data Availability:} \textit{All data underlying the results are available upon request.}
\\


\begin{thebibliography}{99}
		
	\bibitem{challenge} Spalart, Philippe R., and V. Venkatakrishnan. "On the role and challenges of CFD in the aerospace industry." The Aeronautical Journal 120.1223 (2016): 209-232.
	
	\bibitem{CFDML} Vinuesa, Ricardo, and Steven L. Brunton. "Enhancing computational fluid dynamics with machine learning." Nature Computational Science 2.6 (2022): 358-366.	
		
	\bibitem{lag} Lagaris, Isaac E., Aristidis Likas, and Dimitrios I. Fotiadis. "Artificial neural networks for solving ordinary and partial differential equations." IEEE transactions on neural networks 9, no. 5 (1998): 987-1000.
	
	\bibitem{wang} Wang, Lixin, and Jerry M. Mendel. "Structured trainable networks for matrix algebra." In 1990 IJCNN International Joint Conference on Neural Networks, pp. 125-132. IEEE, 1990.
	
	\bibitem{lee} Lee, Hyuk, and In Seok Kang. "Neural algorithm for solving differential equations." Journal of Computational Physics 91, no. 1 (1990): 110-131.
		
	\bibitem{raisi}
	Raissi, M., Perdikaris, P., \& Karniadakis, G. E. (2019).
	Physics-informed neural networks: A deep learning framework for solving forward and inverse problems involving nonlinear partial differential equations.
	\emph{Journal of Computational Physics}, 378, 686-707.
	
%
%
%
%
%
%
	
	
		
	\bibitem{NSF} Jin, Xiaowei, Shengze Cai, Hui Li, and George Em Karniadakis. "NSFnets (Navier-Stokes flow nets): Physics-informed neural networks for the incompressible Navier-Stokes equations." Journal of Computational Physics 426 (2021): 109951.
	
	\bibitem{raisi4} Raissi, Maziar, Alireza Yazdani, and George Em Karniadakis. "Hidden fluid mechanics: Learning velocity and pressure fields from flow visualizations." Science 367, no. 6481 (2020): 1026-1030.
	
	\bibitem{highspeed} Mao, Zhiping, Ameya D. Jagtap, and George Em Karniadakis. "Physics-informed neural networks for high-speed flows." Computer Methods in Applied Mechanics and Engineering 360 (2020): 112789.
	
%
	
	\bibitem{cylinder} Ang, Elijah Hao Wei, Guangjian Wang, and Bing Feng Ng. "Physics-Informed Neural Networks for Low Reynolds Number Flows over Cylinder." Energies 16, no. 12 (2023): 4558.
	
	\bibitem{incom} Rao, C., H. Sun, and Y. Liu, Physics-informed deep learning for incompressible laminar flows.Theoretical and Applied Mechanics Letters, 2020. 10(3): p. 207-212.
	
	\bibitem{PIKAN1} Wang, Yizheng, Jia Sun, Jinshuai Bai, Cosmin Anitescu, Mohammad Sadegh Eshaghi, Xiaoying Zhuang, Timon Rabczuk, and Yinghua Liu. "A physics-informed deep learning framework for solving forward and inverse problems based on Kolmogorov–Arnold Networks." Computer Methods in Applied Mechanics and Engineering 433 (2025): 117518.
	
	
	\bibitem{PIKAN3} Koenig, Benjamin C., Suyong Kim, and Sili Deng. "KAN-ODEs: Kolmogorov–Arnold network ordinary differential equations for learning dynamical systems and hidden physics." Computer Methods in Applied Mechanics and Engineering 432 (2024): 117397.
	
	
%
%
%
%
%
%
%
	
	\bibitem{PIKAN13} Rigas, Spyros, Michalis Papachristou, Theofilos Papadopoulos, Fotios Anagnostopoulos, and Georgios Alexandridis. "Adaptive training of grid-dependent physics-informed kolmogorov-arnold networks." IEEE Access (2024).
	
	\bibitem{PIKAN14} Shuai, Hang, and Fangxing Li. "Physics-informed kolmogorov-arnold networks for power system dynamics." IEEE Open Access Journal of Power and Energy (2025).
	
	\bibitem{phasechange} Patra, Sanjeet, Manish Agrawal, Prasenjit Rath, and Anirban Bhattacharya. "Physics informed neural network-based framework for two-dimensional phase change problems." Computer Physics Communications (2025): 109854.
	
	\bibitem{supersonic} Boya, Sumanth Kumar, and Deepak N. Subramani. "PINTO: Physics-informed transformer neural operator for learning generalized solutions of partial differential equations for any initial and boundary condition." Computer Physics Communications (2025): 109702.
	
	\bibitem{pinto} Zhao, Zhen-tao, Wei Huang, Zhiping Mao, Chao-yang Liu, and Yao-bin Niu. "Physics-informed neural networks for supersonic flow over cones." Computer Physics Communications (2025): 109782.
	
	\bibitem{discret} Zhang, Linying, Wenjun Ma, Qin Lou, and Jun Zhang. "Simulation of rarefied gas flows using physics-informed neural network combined with discrete velocity method." Physics of Fluids 35, no. 7 (2023).
	
	\bibitem{song} Song, Jiahui, Aiguo Xu, Long Miao, Feng Chen, Zhipeng Liu, Lifeng Wang, Ningfei Wang, and Xiao Hou. "Plasma kinetics: Discrete Boltzmann modeling and Richtmyer–Meshkov instability." Physics of Fluids 36, no. 1 (2024).
	
	\bibitem{ANN1} Anthony, Martin, Peter L. Bartlett, and Peter L. Bartlett. Neural network learning: Theoretical foundations. Vol. 9. Cambridge: cambridge university press, 1999.
	
%
	
	\bibitem{PINN1} Karniadakis, George Em, Ioannis G. Kevrekidis, Lu Lu, Paris Perdikaris, Sifan Wang, and Liu Yang. “Physics-Informed Machine Learning.” Nature Reviews Physics, May 24, 2021. https://doi.org/10.1038/s42254-021-00314-5.
	
	\bibitem{datafree} Luo, Huan, and Stephanie German Paal. “A Data‐free, Support Vector Machine‐based Physics‐driven Estimator for Dynamic Response Computation.” Computer-Aided Civil and Infrastructure Engineering, February 15, 2022. https://doi.org/10.1111/mice.12823.
	
	\bibitem{advection} Vadyala, Shashank Reddy, Sai Nethra Betgeri, and Naga Parameshwari Betgeri. "Physics-informed neural network method for solving one-dimensional advection equation using PyTorch." Array 13 (2022): 100110.
	
	\bibitem{2Dadvection1} Fazio, Riccardo, and Salvatore Iacono. "Pseudo-Spectral Methods for Linear Advection and Dispersive Problems." IAENG International Journal of Applied Mathematics 39, no. 1 (2009).
	
	\bibitem{2Dadvection2} Stiernström, Vidar, Lukas Lundgren, Murtazo Nazarov, and Ken Mattsson. "A residual-based artificial viscosity finite difference method for scalar conservation laws." Journal of Computational Physics 430 (2021): 110100.
	
	\bibitem{2Dadvection3} Ali, Ihteram, Sirajul Haq, Kottakkaran Sooppy Nisar, and Shams Ul Arifeen. "Numerical study of 1D and 2D advection-diffusion-reaction equations using Lucas and Fibonacci polynomials." Arabian Journal of Mathematics 10, no. 3 (2021): 513-526.
	
	\bibitem{2Dadvection4} Tsega, Endalew Getnet. "Numerical Solution of Two‐Dimensional Nonlinear Unsteady Advection‐Diffusion‐Reaction Equations with Variable Coefficients." International Journal of Mathematics and Mathematical Sciences 2024, no. 1 (2024): 5541066.
	
	\bibitem{2Dpolar} Sharma, Rishabh Prakash, and Neeraj Kumar. "Nodal integral method for convection-diffusion transport using linear and higher order quadrilateral elements." Numerical Heat Transfer, Part B: Fundamentals 74, no. 3 (2018): 623-645.
	
	\bibitem{curve1} Visbal, Migual, and Datta Gaitonde. "High-order accurate methods for unsteady vortical flows on curvilinear meshes." In 36th AIAA Aerospace Sciences Meeting and Exhibit, p. 131. 1998.
	
	\bibitem{curve2} Visbal, Miguel R., and Datta V. Gaitonde. "On the use of higher-order finite-difference schemes on curvilinear and deforming meshes." Journal of Computational Physics 181, no. 1 (2002): 155-185.
	
	\bibitem{curve3} Hejranfar, Kazem, and Ali Ghaffarian. "A spectral difference lattice Boltzmann method for solution of inviscid compressible flows on structured grids." Computers \& Mathematics with Applications 72, no. 5 (2016): 1341-1368.
	
	\bibitem{Boltzmann1} Kremer, Gilberto M. An introduction to the Boltzmann equation and transport processes in gases. Springer Science \& Business Media, 2010.
	
	\bibitem{Vlasov2} Vlasov, Anatoliĭ Aleksandrovich. "Many-particle theory and its application to plasma." New York (1961).
	
	\bibitem{Vlasov3} Bittencourt, José A. Fundamentals of plasma physics. Springer Science \& Business Media, 2013.
	
%
%
	
	\bibitem{waterbag3} Colombi, S., and J. Touma. “Vlasov-Poisson in 1D: Waterbags.” Monthly Notices of the Royal Astronomical Society, May 26, 2014. https://doi.org/10.1093/mnras/stu739.
	
	
	\bibitem{emfra} Franck, Emmanuel, Ibtissem Labanni, Youssouf Nasseri, Laurent Navoret, Giuseppe Parasiliti Rantone, and Guillaume Steimer. "Hyperbolic reduced model for Vlasov-Poisson equation with Fokker-Planck collision." ESAIM: Proceedings and Surveys 77 (2024): 213-228.
	
	\bibitem{Vlasov4} Lashmore-Davies, C. N. "Kinetic theory and the Vlasov equation.", CERN., (1987).
	
	\bibitem{Vlasov5} Bertrand, Pierre, Daniele Del Sarto, and Alain Ghizzo. The Vlasov Equation 1: History and General Properties. John Wiley \& Sons, 2019.
	
	\bibitem{waterbag2} Besse, Nicolas, Florent Berthelin, Yann Brenier, and Pierre Bertrand. "The multi-water-bag equations for collisionless kinetic modeling." Kinet. Relat. Models 2, no. 1 (2009): 39-80.

	\bibitem{freestream} Manela, A., and L. Gibelli. "Free-molecular and near-free-molecular gas flows over backward facing steps." Journal of Fluid Mechanics 889 (2020): A22.	
	
	\bibitem{relax1} Joyce, Michael, and Tirawut Worrakitpoonpon. "Relaxation to thermal equilibrium in the self-gravitating sheet model." Journal of Statistical Mechanics: Theory and Experiment 2010, no. 10 (2010): P10012.
	
	\bibitem{relax6} Bindoni, Daniele, and L. Secco. "Violent relaxation in phase-space." New Astronomy Reviews 52, no. 1 (2008): 1-18.
	
	\bibitem{Boltzmann2} Cercignani, C. "The Boltzmann equation and fluid dynamics." In Handbook of mathematical fluid dynamics, vol. 1, pp. 1-69. North-Holland, 2002.
	
	\bibitem{Vlasov1} Sinitsyn, Alexander, Eugene Dulov, and Victor Vedenyapin. "Kinetic Boltzmann, Vlasov and related equations." Amsterdam: Elsevier, 2011.
	


\end{thebibliography}

\end{document}